\documentclass[sn-mathphys,Numbered]{sn-jnl}

\pdfoutput=1 
\usepackage{graphicx}%
\usepackage{multirow}%
\usepackage{amsmath,amssymb,amsfonts}%
\usepackage{amsthm}%
\usepackage{mathrsfs}%
\usepackage{xcolor}%
\usepackage{textcomp}%
\usepackage{manyfoot}%
\usepackage{booktabs}%
\usepackage{algorithm}%
\usepackage{algorithmicx}%
\usepackage{algpseudocode}%
\usepackage{listings}%
\usepackage{annotate-equations}%
\usepackage{longtable}
\usepackage{setspace}
\usepackage{caption}
\usepackage{fullpage}
\begin{document}

\title{Evidence for spin-fluctuation-mediated superconductivity in electron-doped cuprates}

\author[1,‡]{\fnm{C. M.} \sur{Duffy}}%
\equalcont{These authors contributed equally to this work.}
\author[2]{\fnm {S. J.} \sur{Tu}}
\equalcont{These authors contributed equally to this work.}
\author[2,3]{\fnm{Q. H.} \sur{Chen}}
\author[2]{\fnm{J. S.} \sur{Zhang}}
\author[1]{\fnm{A.} \sur{Cuoghi}}
\author[6]{\fnm{R. D. H.} \sur{Hinlopen}}
\author[4]{\fnm{T.} \sur{Sarkar}}
\author[4]{\fnm{R. L.} \sur{Greene}}
\author[2,3,5]{\fnm{K.} \sur{Jin}}
\author[1,6]{\fnm{N. E.}\sur{Hussey}}

\affil[1]{\small{\orgdiv{High Field Magnet Laboratory (HFML-FELIX)} and \orgdiv{Institute for Molecules and Materials}, \orgname{Radboud University}, \orgaddress{\city{Nijmegen}, \country{Netherlands}}}}


\affil[2]{\small{\orgdiv{Beijing National Laboratory for Condensed Matter Physics}, \orgname{Institute of Physics}, \orgaddress{\street{Chinese Academy of Sciences}, \city{Beijing}, \country{China}}}}


\affil[3]{\small{\orgdiv{Songshan Lake Materials Laboratory}, \orgaddress{\street{Dongguan}, \city{Guangdong}, \country{China}}}}


\affil[4]{\small{\orgdiv{Maryland Quantum Materials Center and Department of Physics, University of Maryland}, \orgname{College Park}, \orgaddress{\city{Maryland} \postcode{20742}, \country{USA}}}}


\affil[5]{\small{\orgdiv{Key Laboratory of Vacuum Physics}, \orgname{School of Physical Sciences}, \orgaddress{\street{University of Chinese Academy of Sciences}, \city{Beijing}, \country{China}}}}

\affil[6]{\small{\orgdiv{H. H. Wills Physics Laboratory}, \orgname{University of Bristol}, \orgaddress{\city{Bristol}, \country{UK}}}}


\abstract{
\textbf{In conventional, phonon-mediated superconductors, the transition temperature $T_c$ and normal-state scattering rate $1/\tau$ - deduced from the linear-in-temperature resistivity $\rho(T)$ - are linked through the electron-phonon coupling strength $\lambda_{\rm ph}$ \cite{Allen-2000}. In cuprate high-$T_c$ superconductors, no equivalent $\lambda$ has yet been identified, despite the fact that at high doping, $\alpha$ - the low-$T$ $T$-linear coefficient of $\rho(T)$ - also scales with $T_c$ \cite{Cooper-Science-2009, Jin-Nature-2011, Putzke-2021, Yuan-Nature-2022, Harada-2022}. Here, we use dc resistivity and high-field magnetoresistance to extract $\tau^{-1}$ in electron-doped La$_{2-x}$Ce$_x$CuO$_4$ (LCCO) as a function of $x$ from optimal doping to beyond the superconducting dome.  A highly anisotropic inelastic component to $\tau^{-1}$ is revealed whose magnitude diminishes markedly across the doping series. Using known Fermi surface parameters and subsequent modelling of the Hall coefficient, we demonstrate that the form of $\tau^{-1}$ in LCCO is consistent with scattering off commensurate antiferromagnetic spin fluctuations of variable strength $\lambda_{\rm sf}$. The clear correlation between $\alpha$, $\lambda_{\rm sf}$ and $T_c$ then identifies low-energy spin-fluctuations as the primary pairing glue in electron-doped cuprates. The contrasting magnetotransport behaviour in hole-doped cuprates \cite{Konstantinovc-2000, Ayres-2021} suggests that the higher $T_c$ in the latter cannot be attributed solely to an increase in $\lambda_{\rm sf}$. Indeed, the success in modelling LCCO serves to reinforces the notion that resolving the origin of high-temperature superconductivity in hole-doped cuprates may require more than a simple extension of BCS theory.}
}





\maketitle






The BCS theory of superconductivity, wherein the electron-phonon interaction binds two electrons into a Cooper pair, forms the basis of our present understanding of conventional superconductors. The same interaction is also known to scatter electrons in the normal (resistive) state; the associated scattering rate $1/\tau$ determining the $T$-dependence of the electrical resistivity $\rho(T)$. At intermediate temperatures ($\sim \theta_D/4 < T < \theta_D$, where $\theta_D$ is the Debye temperature), $\rho(T)$ is $T$-linear and
$\hbar/\tau = 2 \pi \lambda_{\rm tr} k_B T$. Here $k_B$ is the Boltzmann constant and $\lambda_{\rm tr}$ is a measure of the electron-phonon coupling strength as manifest in the charge transport. Likewise, $T_c$ is governed by an electron-phonon coupling constant $\lambda_{\rm ph}$ through the McMillan formula  \cite{McMillan-PhysRev-1968}. A near equivalence of $\lambda_{\rm ph}$ and $\lambda_{\rm tr}$ has been demonstrated for a wide array of elements and alloys \cite{Allen-2000} (see Fig.~7 and accompanying text in Methods), affirming the correlation between the pairing strength and the strength of the relevant (transport) scattering process. 

Many superconductors of current interest, including heavy fermions, Fe-based superconductors, organic conductors, nickelates and cuprates, are believed to have a pairing mechanism that is non-phononic in origin. A common feature in many of these systems is their proximity to magnetic order, thus motivating models based on coupling to spin fluctuations \cite{Scalapino-1986, Chubukov-2008}. (TMTSF)$_2$PF$_6$, for example, undergoes a spin density wave (SDW) transition with an onset temperature that is suppressed with applied pressure, giving way to a dome of superconductivity that encompasses the critical point at which the SDW phase vanishes. Beyond the SDW ordered phase, a $T$-linear resistivity is observed at low-$T$ whose coefficient $A_1$ scales with $T_c$ \cite{Doiron-Leyraud_PRB_2009}. Weak-coupling theory attributes this correlation to a varying strength of short-range antiferromagnetic (AFM) spin correlations with increasing pressure \cite{Bourbonnais-2011}. To date, however, it has not been possible to demonstrate experimentally a link between the variation of $A_1$ and a reduction in inelastic scattering off such low-energy spin fluctuations.  

A similar correlation between $A_1$ and $T_c$ also exists in overdoped (OD) cuprates \cite{Cooper-Science-2009, Jin-Nature-2011, Putzke-2021, Yuan-Nature-2022, Harada-2022} and by analogy, a common pairing mechanism has been inferred \cite{Jin-Nature-2011}. As with (TMTSF)$_2$PF$_6$, however, no link has yet been established between $A_1$ and a coupling parameter ($\lambda$) of variable strength. Resistivity is, of course, dependent on a number of factors, not simply the magnitude of the scattering rate, and our inability to isolate the form and strength of 1/$\tau$ has led to alternative proposals for the origin of the $A_1T$ term emerging. It has been argued, for instance, that the $T$-linear scattering rate in overdoped cuprates is in fact universal and tied to the so-called Planckian bound ($\hbar/\tau \sim k_B T$) \cite{Legros-2019, Ayres-2023}. In this circumstance, $A_1$ is determined solely by the (inverse) Fermi temperature $T_F$ \cite{Legros-2019} or by the number density of Planckian carriers \cite{Ayres-2023}. Even the very notion of a $T$-linear scattering rate in cuprates has been challenged \cite{Pelc-2020}. As a result, identification of the primary pairing mechanism remains elusive.

In principle, magnetoresistance (MR) measurements can shed light on the origins of the normal state scattering processes since, according to Boltzmann transport theory, the fractional MR is independent of carrier density and is determined largely by the magnitude of $1/\tau$ (extracted from the mean-free-path $\ell$) and its variation around the Fermi surface (FS). In order to shed light on the pairing mechanism, however, it is necessary to demonstrate a correlation between the strength of such scattering and $T_c$. While angle-dependent, interlayer MR measurements (i.e. with $I \| c$) have had some success in isolating the transport scattering rate in two different families of hole-doped cuprates \cite{AbdelJawad-Nature-2006, French_2009, Grissonnanche-Nature-2021}, corresponding attempts to model the in-plane MR response within a Boltzmann approach have so far been met with mixed results \cite{Ayres-2021, Ayres-2023, Hinlopen-PRR-2022, Ataei-NatPhys-2022}. Certainly, no robust connection has yet been made between the in-plane MR response and the nature of the dominant inelastic scattering processes and thus, between the latter and the strength of superconductivity. 


\begin{figure*}
    \centering
    \includegraphics[width = 1\textwidth]{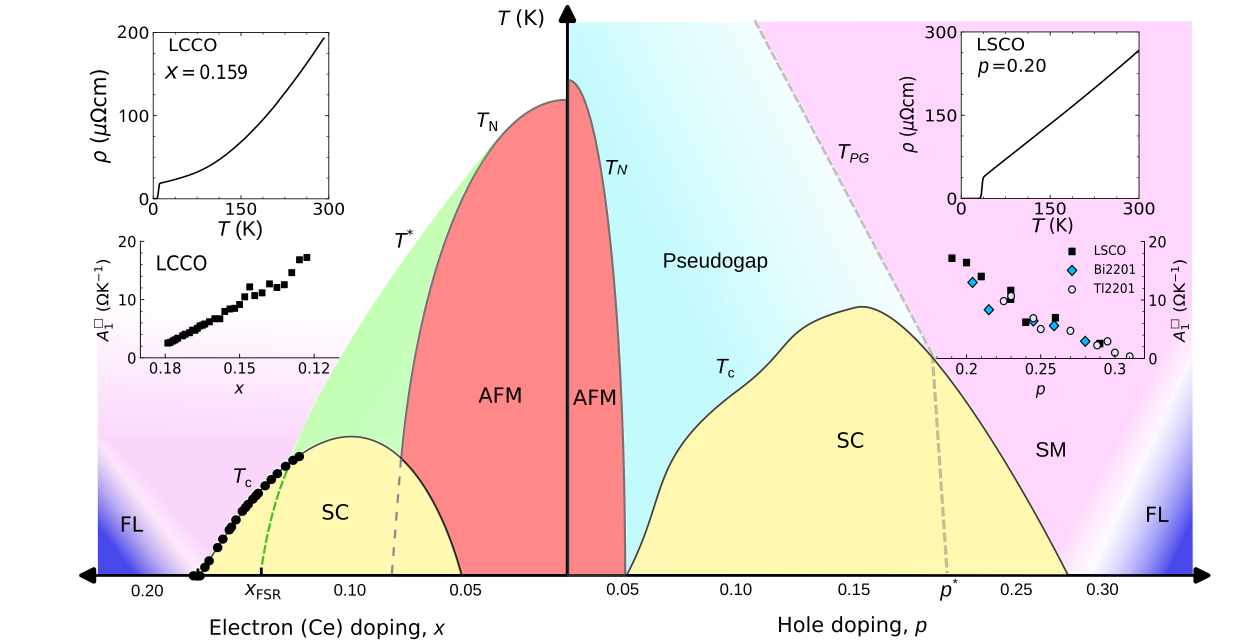}
    \caption{\textbf{Phase diagram of the electron ($n$)- and hole ($p$)- doped cuprates.} Either side of the parent Mott state, the phase diagram of the cuprates exhibits a marked asymmetry. On the $p$-doped side, long-range antiferromagnetic (AFM) order is suppressed prior to the onset of superconductivity (SC) while on the $n$-doped side, AFM order coexists with SC albeit over a narrow doping range. Short range AFM correlations persist into the overdoped side (green dome bounded by $T^{\ast}$), causing a Fermi surface reconstruction that terminates at $x_{\rm FSR}$ \cite{Yu-PRB-2007,Charpentier-PRB-2010}. The $p$-doped cuprates host SC over a more extended range of doping that peaks at a higher maximum transition temperature. They also play host to the pseudogap phase marked in blue, bounded by $T_{PG}$. The strange metal (SM) phase lies beyond the pseudogap phase and is characterised by a predominantly linear-in-$T$ resistivity with a scattering rate corresponding to the Planckian dissipation limit, an example of which is shown in the inset for LSCO at $p=0.20$. By contrast, the resistivity of the overdoped $n$-doped cuprates ($x > x_{\rm FSR}$) is approximately~$T$-linear only at the lowest temperatures \cite{Fournier-1998}, becoming quadratic above $T\sim$ 70~K. The resistivity of LCCO at $x=0.159$ is shown in the upper inset on the left. On both sides of the phase diagram, the coefficient of the $T$-linear resistivity $A_1$ decreases in parallel with $T_c$, as shown in the lower insets for LCCO (from Ref.~\cite{Yuan-Nature-2022}), and for three $p$-doped cuprate families (from Ref.~\cite{Ayres-2023}). Both cuprates host a correlated Fermi liquid (FL) phase in the very overdoped regime.}
    \label{fig:1-PhaseDiagram}
\end{figure*}

In this article, we report a detailed and comprehensive study of the in-plane resistivity and MR of the electron-doped cuprate La$_{2-x}$Ce$_{x}$CuO$_{4}$ (LCCO) in fine doping steps ($\Delta x \lesssim 0.003$) across the entire overdoped region of the phase diagram, from just above optimal $T_c$ ($x$ = 0.116) to the edge of the superconducting dome ($x$ = 0.188). The fine doping steps are achieved by a unique combinatorial film growth technique that synthesizes LCCO films with a continuous composition spread under identical conditions on a single SrTiO$_3$ substrate. More details on the combinatorial technique can be found in Ref.~\cite{Yuan-Nature-2022}. The 19 different doping levels investigated in this study are marked as solid circles in Fig. \ref{fig:1-PhaseDiagram}. The study reveals an orbital MR with distinct temperature and doping dependences that can be captured within a semi-classical Boltzmann treatment by invoking strong anisotropy in the inelastic scattering channel associated with the presence of \lq hotspots'. We describe a method to incorporate the effect of current vertex corrections into the Boltzmann equation and apply this to model the Hall coefficient $R_{\rm H}$ in LCCO, thereby enabling us to identify the location of the hotspots on the AFM Brillouin zone boundary. The key quantity, however, is the component of this anisotropic scattering rate that scales linearly with temperature and whose magnitude is found to drop by an order of magnitude across the studied doping range. Collectively, this analysis provides compelling evidence that the conduction electrons couple most strongly to commensurate low-energy AFM spin fluctuations in the normal state in LCCO. Given the correlation between the strength of this coupling and $T_c$, one can then postulate that superconductivity in LCCO and by extension, in all $n$-doped cuprates, is mediated by AFM spin fluctuations.

In order to frame our subsequent discussion, we reproduce in Figure \ref{fig:1-PhaseDiagram} the phase diagram for electron- and hole-doped cuprates. Two features of immediate interest are the different extents of the AFM phase and the form of the in-plane resistivity $\rho(T)$. In $n$-type materials, long-range AFM order coexists with superconductivity over a part of the superconducting dome, while it is suppressed prior to superconductivity in $p$-doped cuprates \cite{Ando-PRL-2001}. Beyond the long-range AFM phase, $n$-doped cuprates host short-range commensurate AFM correlations that give rise to a Fermi surface reconstruction (FSR) terminating at $x=x_{\rm FSR}$ \cite{Thurston-PRL-1990, Yu-PRB-2007, Charpentier-PRB-2010, He-PNAS-2019}. In the overdoped regime, $\rho(T)$ of both $n$- and $p$-doped cuprates contains a dominant $T$-linear component that persists to low-$T$ over an extended doping range \cite{Cooper-Science-2009, Jin-Nature-2011, Putzke-2021, Yuan-Nature-2022}. The respective correlations between $A_1$ and $T_c$ are illustrated in the bottom insets of both panels. The top insets show that while $\rho(T)$ in $p$-doped cuprates remains quasi-$T$-linear up to high temperatures, $\rho(T)$ in $n$-doped cuprates exhibits a crossover to an anomalous $T^2$-dependence that extends to temperatures way beyond what one would expect for a normal Fermi-liquid \cite{Sarkar-PRB-2018}.

\begin{figure*}
    \centering
    \includegraphics[width=\textwidth]{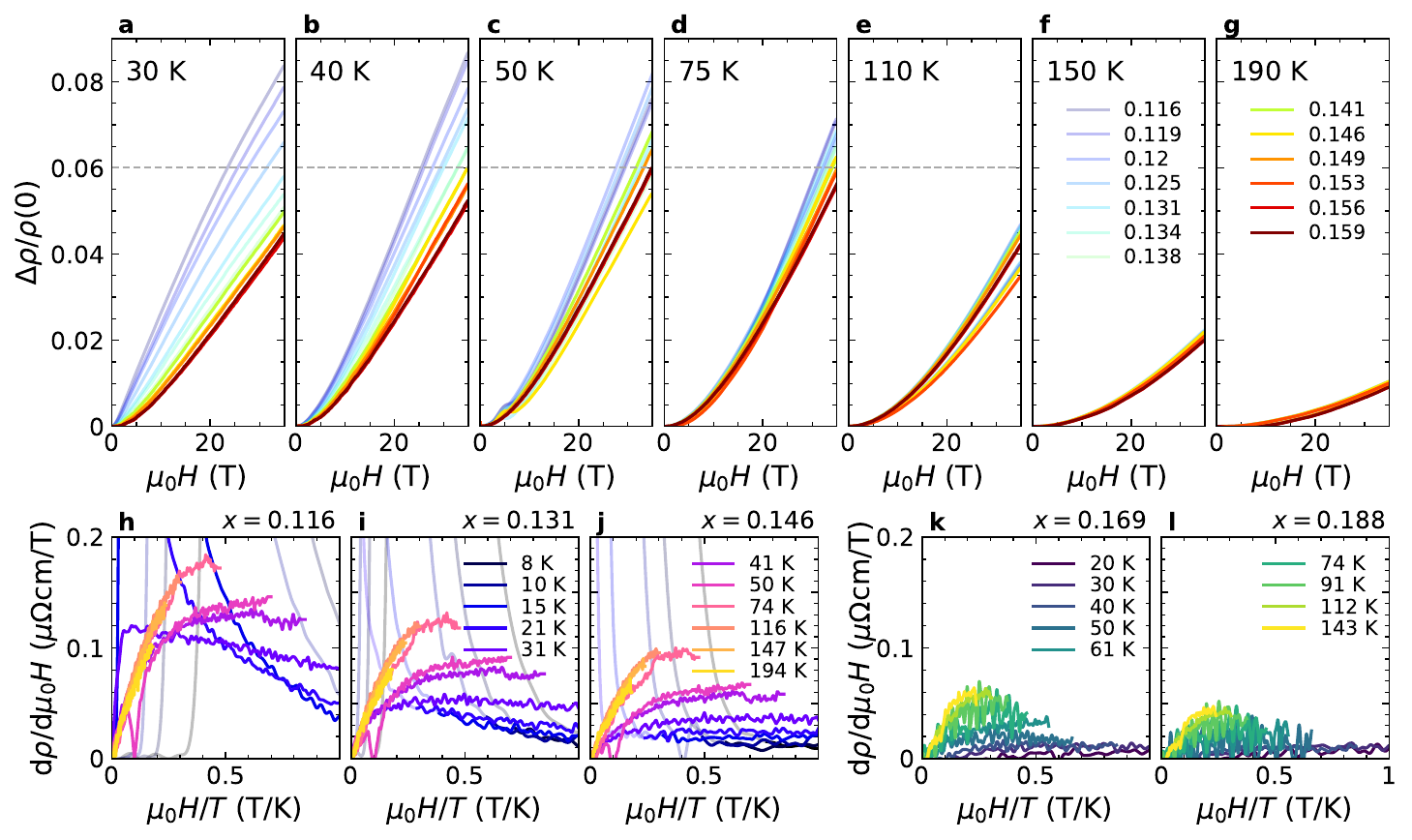}
    \caption{{\bf $T$- and $x$-dependence of the in-plane magnetoresistance in LCCO}. ({\bf a--g}) The MR for 13 channels (0.116 $\leq x \leq$ 0.159) on the same film measured across a broad $T$-range (30 K $\leq T \leq$ 190 K). At the highest temperature, the MR is small and quadratic but as $T$ decreases, the magnitude of the MR grows and becomes more $H$-linear. Surprisingly, for $x>x_{\rm FSR}$, the MR peaks between 50 -- 75 K before falling at lower $T$. Such behaviour -- occurring in the absence of AFM order -- implies that the anisotropy in the mean free path $\ell$ must decrease faster than $\ell$ itself and thus, must exist predominantly in the inelastic scattering channel. See main text for more details. The horizontal dashed grey line marks the peak in the magnitude of the MR for $x=0.159$ at $T$ = 50 K. (\#1.2/1.3 ({\bf h--l}) Plots of d$\rho$/d$(\mu_0H)$ vs.~$H/T$ for 5 different $x$ concentrations (3 from S1, 2 from S2), highlighting the absence of $H/T$ quadrature scaling across the entire doping range. Below 75 K, $H/T$ scaling fails and the slope of the high-field $H$-linear MR decreases. This non-adherence to quadrature scaling is in complete contrast to the $p$-doped cuprates \cite{Ayres-2021}, suggesting that the origin of their $H$-linear MR is fundamentally different. At low-$T$, the $T$-dependence of the MR weakens, but the $H$-linear slope remains. The magnitude of the $H$-linear MR decreases systematically with increasing $x$. Similar results on a limited number of $x > x_{\rm FSR}$ LCCO films were found previously \cite{Sarkar-sciadv-2019}.}
    \label{fig:2-quadrature}
\end{figure*}

Figure \ref{fig:2-quadrature} summarises the most salient features of the in-plane MR response in LCCO. Panels (a)--(g) show isotherms of the MR in 13 channels (0.115 $< x <$ 0.16) measured on a single film (S1). The MR is small at high-$T$, grows with decreasing $T$ and $x$ but then peaks around 50 -- 75 K before decreasing in magnitude at lower $T$. Below $x_{\rm FSR}$ (= 0.14), a negative component can be seen in the MR at high field strengths that has been attributed previously to an isotropic spin effect \cite{Dagan-PRL-2005}. The temperature onset of the negative MR component increases with decreasing $x$ as the long-range AFM phase is approached. 

In panels (h)--(l) of Fig.~\ref{fig:2-quadrature}, we plot d$\rho$/d$\mu_0H$ -- the field derivative of the MR -- as a function of $H/T$ at 5 representative doping levels that span the full studied range. In $p$-doped cuprates, all such derivatives (above the SC state) collapse onto a single curve, a form of scaling dubbed Planckian quadrature scaling \cite{Ayres-2021}. Although the MR in $n$-doped cuprates exhibits a similar $H^2$-to-$H$ crossover with increasing field strength (most evident in panel (i)), the scaling breaks down completely below an $x$-independent temperature of $\sim 75 \ \rm{K}$; the $H$-linear slope at high fields falling monotonically with decreasing $T$ before plateauing (above $x_{\rm FSR}$) or becoming negative (below $x_{\rm FSR}$) below 20 K (Fig.~15 in Methods)). Note that the magnitude of the MR also falls monotonically with increasing $x$ (decreasing $T_c$).

As mentioned above, the MR of a metal is set by $1/\tau$ (or more formally $\ell$) and its variation around the FS. Typically, this anisotropy is smooth and weakly $T$-dependent, resulting in a MR that obeys Kohler scaling ($\Delta \rho \propto f(H/\rho) \propto f(\omega_c\tau))$ $\propto f(\ell))$ where $\omega_c$ is the cyclotron frequency. With decreasing $T$, $\tau$ increases as $\rho$ decreases. Hence, one expects the MR to increase in magnitude as $T$ decreases, as seen in LCCO at elevated temperatures \cite{Poniatowski-PRB-2021}. Below 60 K, however, the MR (for $x > x_{\rm FSR}$) is seen to decrease at a much faster rate than $\rho$ itself decreases. Qualitatively, this observation implies that the size of the MR in LCCO is governed primarily by the degreee of anisotropy in the \textit{inelastic} scattering rate. 

To put this on a more quantitative footing, we model the MR using Boltzmann transport theory. Since the Fermi velocity $v_F$ has only modest ($\leq$ 10\%) angular variation \cite{Horio-NatCom-2016}, we can presume that all $T$-dependent anisotropy in $\ell$ is contained within $\tau^{-1}(\varphi)$ without loss of generality. In order to model $\rho(H, T)$ successfully, the form of $\tau^{-1}(\varphi)$ should reproduce the following three key features of the data:

\begin{enumerate}

\item As shown in Fig.~6e) in Methods, the zero-field resistivity below 60 K can be described accurately by the form $\rho_0 + A_1T + A_2T^2$. With increasing $x$, $A_1$ falls while $A_2$ remains constant, suggesting that only the $T$-linear term is $x$-dependent \cite{Yuan-Nature-2022}.

\item Since the magnitude of the MR falls with decreasing $T$, the anisotropy in $\ell$ must lie predominantly in the inelastic ($T$-dependent) channels.

\item In order to account for the robust $H$-linear MR seen at high field strengths (in some cases up to 65 T \cite{Sarkar-sciadv-2019}) within the same single-fluid Boltzmann picture, we also need to incorporate strong scattering centres (hotspots) into our ansatz that act to impede cyclotron motion \cite{Koshelev-2016, Hinlopen-PRR-2022} and thereby preserve the $H$-linearity of the MR.

\end{enumerate}

\begin{figure}
    \centering
    \includegraphics[scale=0.7]{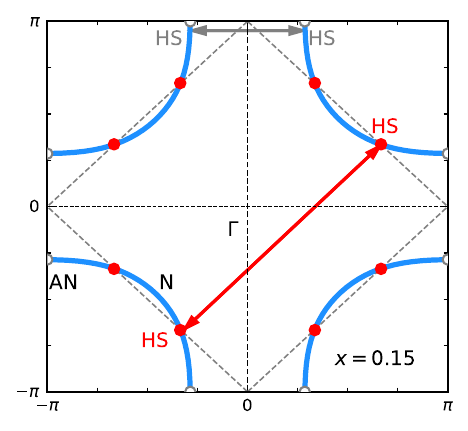}
    \caption{\textbf{The Fermi surface of LCCO ($x=0.15$).} The long double-headed arrow represents large $Q$ (($\pi,\pi$)) hotspot (HS) scattering across opposite sides of the AFM Brillouin zone boundary (marked by the dashed grey line). The short double-headed arrow represents ($\delta$, 0) HS scattering across flat sections of the FS, presumably due to charge correlations \cite{daSilvaNeto-SciAdv-2016}. The locations of the antinodes (AN) and nodes (N) are also marked.}
    \label{fig:3-FS}
\end{figure}

We proceed by applying the Shockley-Chambers Tube Integral Formalism (SCTIF) to the LCCO FS depicted in Figure \ref{fig:3-FS}. (Details of the SCTIF and the FS tight-binding parameters can be found in the Methods section, along with a discussion of the viability of including inelastic scattering processes within a Boltzmann approach.) Treating the MR as an orbital effect is justified by the fact that the longitudinal MR is much smaller than the transverse MR (see Refs.~\cite{Dagan-PRL-2005, Li-SciAdv-2019} and Fig.~8 in Methods). 

Our ansatz for the scattering rate $\tau^{-1}(\varphi, T, x)$ is then constructed as follows. In order to satisfy point 1) above, we introduce 3 terms to represent $\rho(0,T)$: an impurity term $\tau_{\rm imp}^{-1}$ (assumed to be isotropic in line with the quasi-isotropic $v_{\rm F}$ in LCCO), a quadratic term $\alpha_2T^2$ (both $\tau_{\rm imp}^{-1}$ and $\alpha_2$ are fixed throughout) and a doping dependent $T$-linear term $g(x)\alpha_1T$. To account for 3) -- the $H$-linear MR -- we also introduce a \lq hotspot' (HS) term $\tau_{HS}^{-1}(\varphi)$ (a series of sharp Gaussians) at the locations where the AFM zone boundary connects the FS (red arrow in Fig.~\ref{fig:3-FS}), consistent with an early neutron scattering study showing that spin correlations in $n$-doped cuprates remain commensurate at ($\pi, \pi$) within the SC doping range \cite{Yamada_1999}, as well as photoemission experiments that identified  hotspots at the intersection between the FS and the AFM zone boundary \cite{Armitage_PRL_2001, Fujita_PRL_2008, Jenkins-PRB-2010}. \cite{Yamada_1999} \cite{Armitage_PRL_2001, Fujita_PRL_2008, Jenkins-PRB-2010}. As discussed below and presented in Figures 11--13 in Methods, this choice of HS location is corroborated by analysis of the unusual $T$- and $x-$dependence of $R_{\rm H}$ in LCCO \cite{Yuan-Nature-2022, Sarkar-PRB-2017}.

While these four components can account for $\rho(0,T)$ and the crossover to $H$-linear MR at high fields, they fail to capture the unusual increase in the MR with increasing $T$ shown in Fig.~\ref{fig:2-quadrature} (recall that $\tau_{HS}^{-1}(\varphi)$ has no intrinsic $T$-dependence). In addressing point 2), we must therefore incorporate $\varphi$-dependence into one or both of the $T$-dependent terms of $\rho(0,T)$, i.e.~$g(x)\alpha_1T$ and $\alpha_2T^2$ such that as $\tau^{-1}(T)$ increases, so too does its degree of anisotropy. In the following, we assign it to both for simplicity. (\#X.1) Given that $v_{\rm F}(\varphi)$ is isotropic, the most obvious and intuitive source of this anisotropy is the HS itself which we parameterise using a sin$^\nu(2\varphi)$ term, whose exponent $\nu$ serves as a measure of the effective broadening of each HS (which in turn quantifies the inverse correlation length of the ($\pi$, $\pi$) spin fluctuations).

The above considerations define the minimal ingredients of $\tau^{-1}(\varphi, T, x)$ required to reproduce the dc resistivity and magnetotransport properties of LCCO:
\vspace{.3cm}
\begin{equation}
\label{eq:LCCO_scattering}
 \tau^{-1}(\varphi, T, x) =
    \overbrace{\eqnmarkbox[orange]{p1}{\tau_{\rm imp}^{-1} + \tau_{\rm HS}^{-1}(\varphi)}}^{T-\rm{independent}} +
    \overbrace{\eqnmarkbox[blue]{p2}{(g(x)\alpha_1T + \alpha_2 T^2)\sin^{\nu}\left[2\left(\varphi \pm \left(\frac{\pi}{4}- HS\right)\right)\right]}}^{T-\rm{dependent}}
\end{equation}

\begin{figure*}
    \centering
    \includegraphics[width = 1\textwidth]{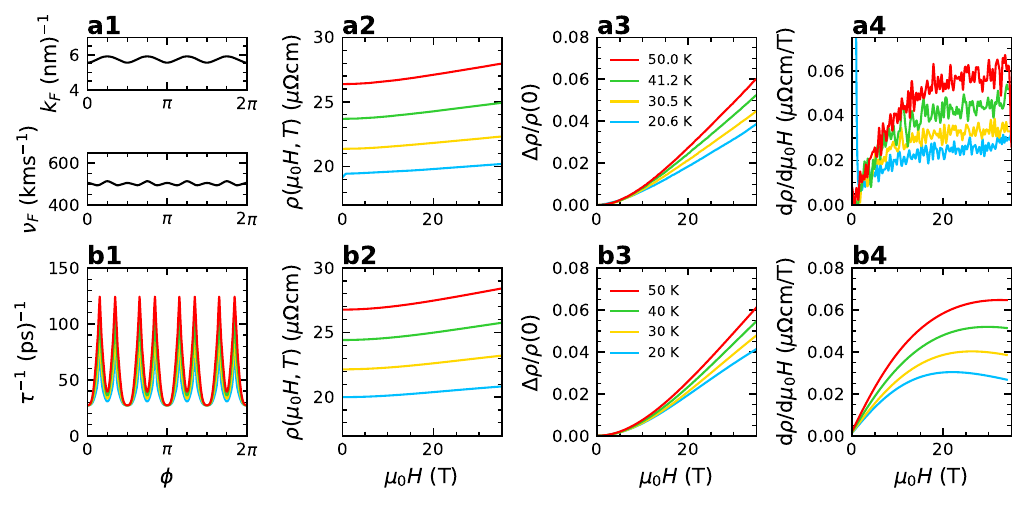}
    \caption{{\bf Optimised simulations of the magnetoresistance for $x=0.159$ (20 K $\leq T \leq$ 50 K)}. \textbf{a1}) Variation of the FS (upper panel) and Fermi velocity $\nu_F$ obtained from the tight-binding parameterisation given in Table 1 in Methods. \textbf{a2-4}) Experimental MR curves, $\Delta\rho/\rho(0)$ and d$\rho$/d($\mu_0H$), respectively, for $x=0.159$. The notable features of the MR are a magnitude that increases with increasing $T$, indicative of developing anisotropy in the inelastic scattering channel, and a $H^2$-to-$H$ crossover with a clear turnover scale most evident in panel a4). \textbf{b1-4}). Simulations resulting from the form of scattering rate expressed in Eq.~(1) and depicted in \textbf{b1}) where the inelastic component is peaked at the hotspots located at the AFM BZ boundary. Here, $HS$ = 27.5$^o$. This scattering rate captures all the main features of the experimental data, as shown in panels \textbf{b2-4}. Alternative scattering rate scenarios are considered in Fig.~9 in Methods. 
    } 
    \label{fig:4-sims}
\end{figure*}

The choice of $\nu$ is somewhat arbitrary, so here, we borrow from a previous parameterisation of the scattering rate in hole-doped cuprates \cite{Grissonnanche-Nature-2021} and set $\nu$ = 12. Again, this value is fixed for all subsequent simulations. A summary of the SCTIF simulations for $x$ = 0.159 (a doping level common to the two films studied in this work and far from the AFM phase) is presented in Figure \ref{fig:4-sims} (see Fig.~10 in Methods for simulations of the MR of other doping levels). Panel a1) of Fig.~\ref{fig:4-sims} illustrates the variation of $k_{\rm F}(\varphi)$ and $v_{\rm F}(\varphi)$ derived from the tight-binding parameterisation of the FS \cite{Tang_PRB_2021}, while panels a2), a3) and a4) show, respectively, the experimental MR curves, the fractional MR and d$\rho$/d$\mu_0H$ between 20~K and {50~K}. Panel b1) depicts the form of $\tau^{-1}(\varphi)$ defined in Eq.~(1) that generates the best simulations of the experimental data -- shown in the remaining panels of row (b). Notably, the same form of $\tau^{-1}(\varphi,T,x)$ also reproduces the magnitude of the MR. Other forms of anisotropy which fail to account for all aspects of the MR are explored in Fig.~9 in Methods.

In order to examine the doping dependence $\tau^{-1}(x)$, we fit the zero-field resistivity and the magnitude of the MR at $T$ = 40 K for all ($\approx$ 13) voltage pairs measured on film S1, using Eq.~(1) and only one doping-dependent variable parameter -- $g(x)$.  (\#1.6) Here, we have assumed that there is no novel state induced by the field which would induce a FS reconstruction or additional components in the scattering rate. (\#1.3) Moreover, while the MR curves for S2 follow the same monotonic trends as in S1 -- see panels k) and l) of Fig.~2 -- the presence of a small negative component in the low-field MR of S2 complicates the analysis of the MR in S2. For this reason, an estimate of $g(x)$ for S2 could only be deduced from the doping dependence of the $T$-linear resistivity coefficient (Fig.~6e) in Methods), having first normalized the value of $g(x)$ at $x$ = 0.159 to the value deduced for S1 at the same doping level.


\begin{figure*}
    \centering
    \includegraphics[width = 1\textwidth]{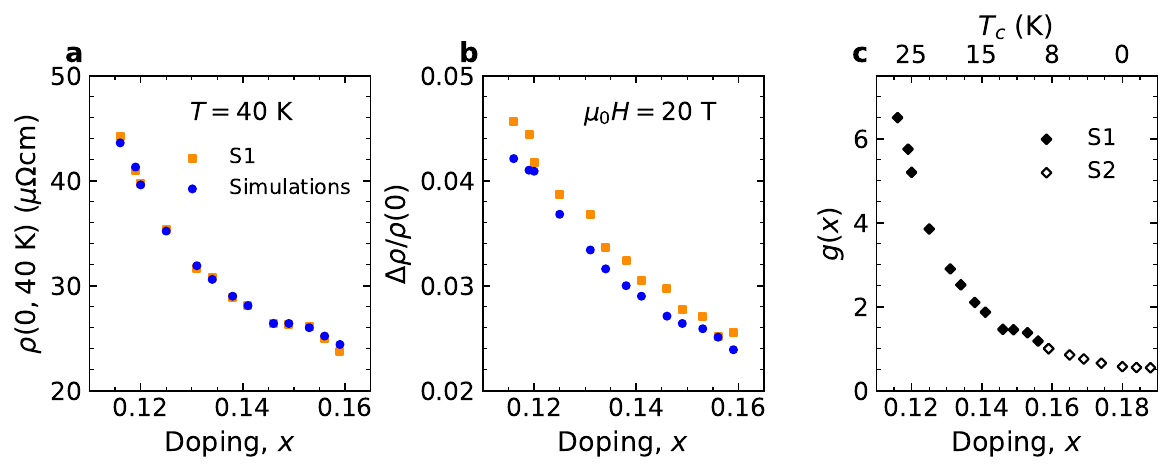}
        \caption{{\bf Coupling parameter for high-$T_c$ superconductivity in electron-doped cuprates}. \\ \textbf{a}) Doping dependent simulations (blue circles) of $\rho$(0, 40 K) -- the zero-field resistivity at $T=40$ K -- where signatures of AFM order are absent or weak across (\#1.3) the doping range of film S1. The orange squares represent the experimental data; the closed blue circles the simulations. \textbf{b}) Corresponding simulations (blue circles) of the fractional MR $\Delta \rho/\rho(0)$ at $\mu_0H$ = 20 T and $T$ = 40~K as a function of $x$. \textbf{c}) The scaling factor $g(x)$ of the anisotropic $T$-linear scattering rate as a function of $x$. Note that $g(x)$ was the only doping-dependent variable parameter used to obtain the results in panels a) and b). Note, too, that for S2, we take $g(x)$ from the coefficient of the $T$-linear resistivity, and normalise it to the value obtained in S1 at $x=0.159$. Across the doping series, $g(x)$ drops by one order of magnitude as $T_c\rightarrow 0$. Thus, $g(x)$ directly relates the anisotropic $T$-linear scattering rate to $T_c$, acting as an effective coupling parameter for both transport and superconductivity.}
    \label{fig:5-slope}
\end{figure*}

The results are summarised in Fig.~\ref{fig:5-slope} and the corresponding fitting parameters and $g(x)$ values are listed in Tables 3 and 4, respectively, in Methods. Panels a) and b) of Fig.~\ref{fig:5-slope} show, respectively, the 3-fold drop in $\rho$(0, 40 K) and the 2-fold drop in $\Delta\rho/\rho(0)$ (at 20 T) across S1. For the MR to obey Kohler scaling as a function of $x$, one expects $\Delta\rho/\rho(0)$ to increase by a similar amount to the decrease in $\rho(0)$. The counter-trend in $\Delta\rho/\rho(0)$ shown in Fig.~\ref{fig:5-slope}(b), however, inevitably requires an even greater decrease in $g(x)$, as revealed in panel c). Coupled with the results from S2, the order-of-magnitude drop in $g(x)$ across the series mirrors the variation in $A_1$ (the coefficient of the $T$-linear resistivity) reported by Yuan \textit{et al.} \cite{Yuan-Nature-2022} and shown in Fig.~\ref{fig:1-PhaseDiagram}. Over the same doping range, $T_c$ is reduced from near its optimal value (= 25 K) to zero. This clear correlation between $T_c$ and the strength of the anisotropic $T$-linear scattering rate is the first key result of this study, revealing as it does the effective coupling parameter for superconductivity in LCCO. These considerations also reveal why the MR is so constraining; (i) in isolating the contribution of $\ell$ to $\rho(T)$, (ii) in revealing the presence of marked anisotropy in $\tau^{-1}(\varphi)$ (as well as the presence of hot-spots through the observation of a robust $H$-linear MR) and (iii) in revealing the marked drop in the strength of this anisotropic scattering across the doping series (and with it, the drop in the effective coupling strength $\lambda$).
 
This brings us to the second key finding of this study, namely the elucidation of the source of this anisotropic scattering, for which we turn to the Hall response of LCCO. In a layered system, the Hall conductivity $\sigma_{xy}$ is determined by the \lq Stokes' area swept out by the $\vec{\ell}(\varphi)$ vector as $\varphi$ varies around the FS \cite{Ong-1991}. The FS of LCCO, as sketched in Fig.~\ref{fig:3-FS}/\ref{fig:4-sims}, has singular curvature, meaning that $\vec{v}_{\rm F}$ has the same circulation at all $\varphi$. In this circumstance, even the inclusion of strong anisotropy in $\tau^{-1}(\varphi)$ cannot cause $\sigma_{xy}$ and thus $R_{\rm H}$ to change sign. In LCCO for $x > x_{\rm FSR}$, however, $R_{\rm H}(T)$ drops monotonically from close to its isotropic (Drude) value towards zero and becomes negative at elevated $T$ (Fig.~11 in Methods \cite{Sarkar-PRB-2017, Yuan-Nature-2022}). In the absence of any additional FSR, for which there is no evidence, the only effective way to explain the sign change in $R_{\rm H}(T)$ is a breakdown of the relaxation time approximation (RTA) and the incorporation of current vertex corrections (CVCs) into the calculation for $\sigma_{xy}$ \cite{Jenkins-PRB-2010}.

In nearly AFM metals with dominant scattering vector $\vec{Q}$ = ($\pi$, $\pi$), the total current $\vec{J}_k$ at the hotspots is approximately parallel to $\vec{v}_k + \vec{v}_{k+Q}$ and is thus no longer perpendicular to the FS. Applying linear response theory, Kontani derived the following expression to describe how $\vec{J}_k$ varies in proximity to the AFM hotspot \cite{Kontani-RepProgPhys-2008}:

   \begin{equation}
        \label{eq:HS_CVC}
            \vec{J}_k = \frac{1}{1 - \epsilon_k^2} (\vec{v}_k + \epsilon_k \vec{v}_{k\pm Q})
    \end{equation}

where $\epsilon_k \approx$ (1 - $c \xi_{\rm AFM}^2) < 1$, $\xi_{\rm AFM}$ is the AFM correlation length and $c$ is a constant. In order to incorporate the effect of CVCs into our Boltzmann analysis, we substitute this expression for $\vec{J}_k$ (i.e. $\vec{v}_k$) into the kinetic equation and recalculate $\sigma_{xx}$, $\sigma_{xy}$ and $R_{\rm H}$ (Fig.~12 in Methods). Without prior knowledge of $\xi_{\rm AFM}$ in LCCO, it is only possible to model $R_{\rm H}$ semi-quantitatively. Nevertheless, as shown in Fig.~13 in Methods, using the same form for $\tau^{-1}(\varphi)$ that was used to model the MR and a reasonable choice for $\epsilon_k$, we are able to capture the key features of  $R_{\rm H}(T, x)$ beyond $x > x_{\rm FSR}$, including the tendency of $R_{\rm H}$ to go negative with increasing $T$. Note that an earlier study of the frequency ($\omega$) dependence of the Hall angle in overdoped Pr$_{2-x}$Ce$_x$CuO$_4$ also interpreted the sign change of $R_{\rm H}(T, \omega)$ as evidence for AFM CVCs in a FS with singular (hole-like) curvature \cite{Jenkins-PRB-2010}. Moreover, CVCs arising from AFM spin fluctuations have been shown to generate a $T + T^2$ resistivity consistent with what is found in LCCO \cite{Kontani-RepProgPhys-2008, Bergeron-2011}. Conversely, the other chief candidate for HS scattering, namely that due to charge fluctuations across flat sections of FS \cite{Ishii-natcomm-2014, daSilvaNeto-Science-2015, daSilvaNeto-SciAdv-2016, Hepting-Nature-2018} (horizontal double-headed arrow in Fig.~3) is found to give rise to a $T$-dependence of $R_{\rm H}(T)$ that is counter to what is observed experimentally (see Fig.~13 in Methods).

Further support for dominant scattering off AFM spin fluctuations is provided through inspection of the high-field slope of the MR over the full ($T$, $x$) range (Fig.~15 in Methods). The low-$T$ negative contribution to the MR seen at low $x$ and attributed to long-range AFM order \cite{Dagan-PRL-2005} crosses over smoothly to positive MR with no discontinuity near $x_{\rm FSR}$, suggesting that the two components share the same origin. This combined analysis of the Hall and MR responses enables us to identify ($\pi$, $\pi$) scattering as the dominant normal-state scattering in LCCO. Taken together with the observed correlation between $g(x)$ and $T_c$, we conclude that SC pairing in $n$-doped cuprates is mediated by the exchange of low-energy spin fluctuations.

Before discussing some of the other implications of our findings, we first highlight two aspects of the magnetotransport in LCCO that are not yet understood. First among these is the $T^2$ resistivity that extends to high $T$ \cite{Bach-PRB-2011, Sarkar-PRB-2018}. As shown in Fig.~\ref{fig:1-PhaseDiagram}, this behaviour contrasts markedly with the $T$-linearity found in $p$-doped cuprates at high-$T$. The fact that $\rho_{ab}(T)$ in LCCO extends beyond the estimated Mott-Ioffe-Regel limit suggests that this super-linear behaviour may arise from a gradual loss of carriers or a transfer of spectral weight with increasing $T$. The latter could be revealed, for example, in optical conductivity measurements at elevated temperatures \cite{Hussey-2004}. The other outstanding issue is the persistence of $H$-linear MR to the lowest $T$ \cite{Sarkar-sciadv-2019}. Within the model presented above, the $H$-linear MR is a result of impeded cyclotron motion due to the presence of hotspots. Of course, there may be a physical reason for this, e.g. the presence of a quantum critical phase and associated residual scattering off quantum fluctuations associated with the AFM order, but it is not immediately obvious why it should persist down to low-$T$. An alternative explanation is that this residual, small $H$-linear MR arises due to the compositional variation along the film created by the combinatorial synthesis itself. Longitudinal variation in the Hall voltage is a longstanding explanation for $H$-linear MR in metals \cite{Bruls-1981} and indeed, has recently been proposed in the context of the cuprates \cite{Singleton-2020}. We note, however, that an identical $H$-linear MR is also observed in LCCO films that are not grown combinatorially \cite{Sarkar-sciadv-2019}.

Returning to the implications of our study, there are four additional key points to make. Firstly, it has been argued previously that the magnitude of the $T$-linear resistivity in cuprates is consistent with the notion of a Planckian bound on scattering \cite{Legros-2019}. The analysis presented here, however, reveals that the magnitude of the scattering rate in LCCO varies by almost one order of magnitude across the doping series. Hence, the low-$T$ $T$-linear resistivity in $n$-doped cuprates cannot be tied to the Planckian limit. Our study also reveals that the $T$-linear scattering rate in cuprates is not always isotropic \cite{AbdelJawad-Nature-2006, Grissonnanche-Nature-2021}. Secondly, the persistence of the $T$-linear term in $\rho(T)$ down to low $T$ supports the arguments of Rosch that in the presence of impurity scattering, any shorting effects arising from the \lq colder' regions of the FS away from the hotspots are diminished \cite{Rosch-1999}. Thirdly, given the similarities between the phase diagrams of $n$-doped cuprates and the Bechgaard salts and the scaling between $A_1$ and $T_c$ observed in both families \cite{Yuan-Nature-2022, Doiron-Leyraud_PRB_2009}, it is perhaps timely to investigate the MR response of the latter in order to elucidate $\tau^{-1}(\varphi, T)$ and to determine whether an equivalent coupling parameter could also be identified there.

This brings us finally to the dichotomy between the electron-doped cuprates and their hole-doped counterparts. The asymmetry in their respective phase diagrams (Fig.~\ref{fig:1-PhaseDiagram}) has often led to the assertion that doping the parent Mott insulator with electrons or holes leads to markedly different low-energy physics. Certainly, the $p$-type cuprates are more strongly correlated \cite{Weber-NatPhys-2010}, lying within the strong coupling regime where $U/t$ is about twice as large as in the $n$-doped cuprates ($U$ is the on-site Coulomb repulsion and $t$ is the nearest-neighbour hopping parameter). The ratio $t/J$, where $J=4t^2/U$ is the AFM exchange interaction, is also larger in the $p$-doped cuprates, indicating a lower propensity to remain antiferromagnetic \cite{Macridin-PRB-2005}.

Given these differences, it seems quite remarkable that the low-$T$ resistivities in both wings exhibit such similar behaviour, namely $\rho_{ab}(T)$ = $\rho_0$ + $A_1T$ + $O(T^2)$ with a coefficient $A_1$ that scales with $T_c$. Our present study has shown that such scaling in LCCO reflects directly the coupling to low-energy AFM spin fluctuations and that weak-coupling calculations exploring this link \cite{Moriya-2000, Bourbonnais-2011, Wu-PNAS-2022} may well reach consensus with experiment if applied to the electron-doped side. It is naturally tempting, therefore, to speculate that the same correspondence is at play in $p$-doped cuprates and that the higher-$T_c$ values in the latter are nothing more than a consequence of a larger $\lambda_{\rm sf}$ or comparable $\lambda_{\rm sf}$ coupled with larger $U$. If that were the case, one would expect the MR response of overdoped $p$-type cuprates to exhibit qualitatively similar behaviour, i.e. governed by an anisotropic inelastic scattering rate whose magnitude would drop with decreasing $T$. Yet across the entire strange metal regime in $p$-doped cuprates, the MR response is fundamentally different, in the sense that the MR exhibits a form of quadrature $H/T$ scaling that no variant of kinetic quasiparticle theory has yet been able to account for \cite{Ayres-2021, Ayres-2023}. Instead, there is growing experimental evidence that the strange metallic state on the hole-doped side has a dual character, playing host to both coherent (Boltzmann-like) and incoherent (non-Boltzmann-like) carriers \cite{Ayres-2021, Ayres-2023,Yang-Nature-2022}. Turning this around, one might argue that the inability to apply Boltzmann theory to the hole-doped strange metal is simply a reflection of the fact that the correct parameterisation has not yet been found. Its wholesale applicability to LCCO, however, provides a strong counter-argument that such a parameterisation should already have been found, were it to exist. Indeed, this inability to extrapolate our findings in LCCO to the hole-doped side gets to the heart of the problem (the difference between the electron- and hole-doped cuprates) and implies that the higher $T_c$ values in the latter cannot be simply due to an enhanced coupling to spin fluctuations. This also suggests that an understanding of both the strange metallicity and (high-temperature) superconductivity in $p$-doped cuprates is likely to require a fundamentally new approach, one that lies beyond the quasiparticle picture \cite{Phillips-Science-2022,Patel-Science-2023} or an extension of BCS theory to incorporate spin-fluctuation-mediated pairing. For the latter, at least, it would appear that electron-doped cuprates represent its epitome.


\newpage
\section*{Material and experimental procedures}\label{methods}

Combinatorial La$_{2-x}$Ce$_x$CuO$_4$ (LCCO) thin films were fabricated in Beijing using molecular beam epitaxy from targets with $x$ = 0.10 and 0.19, facilitating measurements over a large region of the LCCO phase diagram in minute steps in $x$. A comprehensive description of the deposition and patterning processes can be found elsewhere \cite{Yuan-Nature-2022}. Two films, S1 (0.116 $\leq x \leq$ 0.159) and S2 (0.159 $\leq x \leq$ 0.188), were patterned into microbridges and their in-plane resistivity measured using a standard four-point ac lock-in method. Seven channels could be measured simultaneously on a single film and in total, 13 were measured on S1 and 7 on S2 with one common doping level at $x$ = 0.159. As shown in Fig.~6a), the two $\rho(T)$ curves at $x$ = 0.159 coincide below $T$ = 70 K provided that the data for S2 are scaled (reduced) by a factor of 1.5. This same scaling factor was then applied to all channels in S2.

Panel b) of Fig.~6 shows a comparison of the $\rho(T)$ curves from film S1 with previous data reported in Ref.~\cite{Yuan-Nature-2022} at similar doping levels. We find that the overall resistivity values in the new films are approximately 1/3 of those reported in Ref.~\cite{Yuan-Nature-2022} and their residual resistivity ratios are also higher. There are two probable reasons for these observations. Firstly, in Ref.~\cite{Yuan-Nature-2022}, the film thickness was estimated from the fabrication conditions, likely leading to deviations from the real value. In the present study, a step profiler was used to measure the thickness, so in principle the as-determined thickness should be closer to the actual value. (We note that a similar multiplicative factor appears in the comparison of the Hall coefficients shown in Fig.~11.) These variations in the absolute magnitudes of $\rho(T)$ do not affect the analysis in Ref.~\cite{Yuan-Nature-2022} since all the reported data were obtained on a single film with the same nominal thickness. Secondly, continual improvement in the quality of the combinatorial films has led to larger residual resistivity ratios and slightly higher $T_c$ values, possibly reflecting improved crystallinity or a more homogeneous oxygen content. Crucially, when the data are scaled such that their $\rho(T)$ curves have the same slope and then adjusted vertically to account for the differences in their residual resistivities, as shown in Fig.~6c), the curves overlap.

Note that the data shown in panels b) and c) of Fig.~6 for film S1 are only discrete points (the zero-field values for each field sweep) rather than a continuous curve. To obtain a full, uninterrupted $\rho(0,T)$ curve between 2 K and 290 K inside the Bitter magnet would have required at least 8 hours of continuous measurement, time that was not available at the high-field facility when measuring S1. Fortunately for S2, the cryostat was available for an extended period following the measurement, making it possible to perform a warming run in zero-field over three days.

\begin{figure}
    \centering
    \includegraphics[width=0.9\textwidth]{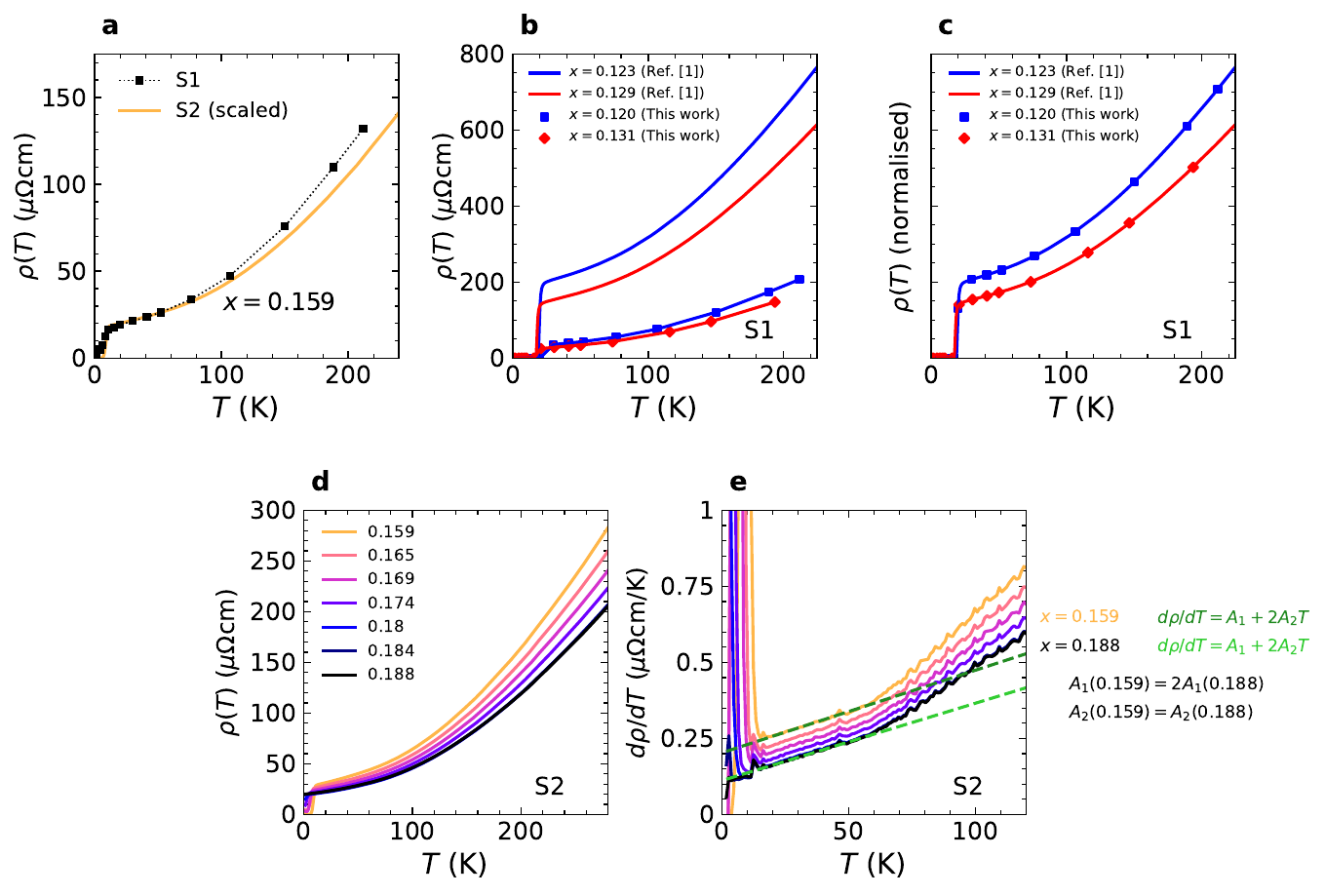}
    \caption{\textbf{Zero-field resistivity of overdoped LCCO films}. (\textbf{a}) $\rho(0,T)$ at $x=0.159$ in the two films. Note that the resistivity values of S2 have been divided by 1.5 in order to match those of S1 below 70~K. (\textbf{b}) As-measured $\rho(0,T)$ data on S1 (symbols) with data for comparable $x$ values from Ref.~\cite{Yuan-Nature-2022}. (\textbf{c}) Closer comparison of the two data sets having scaled the slopes of each curve in Ref.~\cite{Yuan-Nature-2022} and shifted them vertically to account for differences in their residual resistivities. (\textbf{d}) Representative $\rho(0,T)$ curves for the LCCO film S2. The form of $\rho(0,T)$ is the same across all channels. (\textbf{e}) Corresponding derivatives up to 125 K. The finite intercept as $T\rightarrow 0$ indicates a persistent $T$-linear component whose strength diminishes with $x$. Between 15 K and 60 K, the derivatives reveal a constant $T^2$ component of $\rho(0,T)$ across the doping range. The discontinuous change in the derivative of the resistivity at $T\sim 60$ K \cite{Sarkar-PRB-2018} coincides with the temperature at which the MR and the value of the $H$-linear slope are maximal.}
    \label{SI:ZeroField}
\end{figure}

Panels (d) and (e) of Fig.~\ref{SI:ZeroField} show, respectively, the zero-field in-plane resistivity $\rho(0,T)$ and corresponding derivatives d$\rho$/d$T$ for a selection of channels on S2. As is clear from panel e), $\rho(0,T) \propto A_1T + A_2T^2$ at low-$T$, with $A_2$ constant and $A_1$ decreasing with increasing $x$ (and decreasing $T_c$ \cite{Yuan-Nature-2022}). This is the main reason why the coupling strength $g(x)$ was assigned uniquely to the $T$-linear term in Eq. (1) of the main text. Consistent with previous reports \cite{Bach-PRB-2011, Sarkar-PRB-2018}, $\rho(0,T)$ develops a stronger $T$-dependence at higher $T \ (>70 \ {\rm K})$. In-plane magnetoresistance (MR) measurements were carried out in a 35~T Bitter magnet at HFML-FELIX in Nijmegen with the magnetic field applied perpendicular or parallel to the CuO$_2$ planes. The sample stage was glued with GE varnish to maintain its orientation. 

\section*{Link between $T$-linear scattering rate and $T_c$ in metals}

To highlight the link between the strength of the $T$-linear scattering rate and $T_c$ in ordinary metals, we focus here on the work of Allen \cite{Allen-2000} in which a near-equivalence of the coupling strengths deduced from $T_c$ ($\lambda_{ph}$) and from the coefficient of the $T$-linear resistivity $\alpha$ ($\lambda_{tr}$) was demonstrated for a wide array of elements and alloys. For most elements considered in Allen's treatise, $T_c$ was derived directly from the McMillan formula:

\begin{equation}
    \label{Allen}
    T_c = \frac{\theta_{\rm D}}{1.45} exp\frac{-(1.04(1 + \lambda_{\rm McM}))}{(\lambda_{\rm McM} – \mu^*(1 + 0.62 \lambda_{\rm McM}))}
\end{equation}

\begin{figure}
    \centering
    \includegraphics[width=0.45\textwidth]{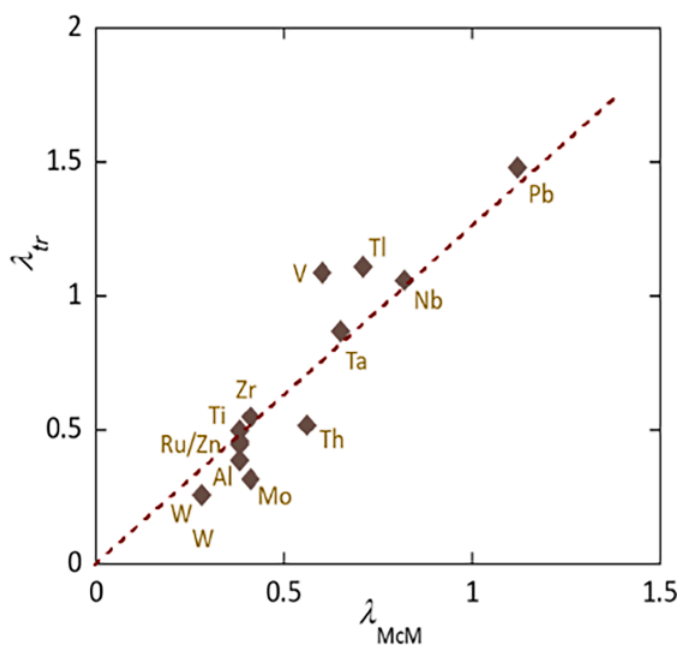}
    \caption{\textbf{$T$-linear resistivity and $T_c$ in conventional metals.} Comparison of $\lambda_{tr}$ – the coupling strength deduced from the $T$-linear resistivity and $\lambda_{\rm McM}$ - the e-ph coupling strength deduced from the value of $T_c$ in a subset of elemental superconductors using the McMillan formula (taken from Ref.~\cite{Hussey-2023}].}
    \label{SI:Allen}
\end{figure}

assuming the same value for the pseudopotential $\mu^*$ consistent with tunneling and theory. Hence, extraction of $\lambda_{\rm McM}$ required only knowledge of $T_c$ and the Debye temperature $\theta_{\rm D}$. For $\lambda_{tr}$, Allen used density functional theory to determine the FS and density of states for the individual elements and then used the Drude equation to deduce the strength of the scattering rate $\Gamma_{tr}$ from $\alpha_1$ – the coefficient of the $T$-linear resistivity. Once this was obtained, Allen determined $\lambda_{tr}$ directly from the expression $\hbar \Gamma_{tr}$ = 2$\pi \lambda_{tr} k_{\rm B} T$. Fig.~\ref{SI:Allen} shows a comparison of $\lambda_{tr}$ and $\lambda_{\rm McM}$ for some elemental superconductors. Despite some scatter around the dotted line in Fig.~\ref{SI:Allen}, the overall trend is clear, indicating a robust correlation between $\lambda_{tr}$ and $\lambda_{\rm McM}$, and through this, between $T_c$ and $\alpha_1$ for almost one decade is variation in $\lambda_{tr}$ ($\lambda_{\rm McM}$). Hence, while there are undoubtedly complicating circumstances, the electron-phonon coupling strength $\lambda_{e-ph}$ does indeed appear to play the decisive role in determining $T_c$ and $\alpha_1$.

\section*{Longitudinal magnetoresistance}

\begin{figure}
    \centering
    \includegraphics[scale=0.9]{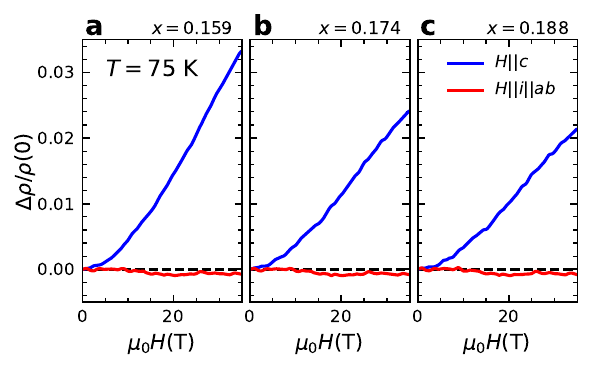}
    \caption{{\bf Orientation dependence of the MR}. The MR of LCCO with $H||c$ (blue) and $H||i||ab$ (red) for three channels ({\bf a}) $x=0.159$, ({\bf b}) $x=0.174$ and ({\bf c}) $x=0.188$ at $T = 75$ K. The MR is anisotropic with respect to field orientation and is close to zero in the parallel configuration as is expected from conventional theories of transport. }
    \label{SI:Hparallel}
\end{figure}

In a quasi-two-dimensional (quasi-2D) metal hosting Landau quasiparticles (QPs), a magnetic field oriented parallel to the direction of the injected current will generate a negligible Lorentz force and as a result, the orbital MR is expected to be close to zero. In the $p$-doped cuprates Bi$_2$Sr$_2$CuO$_{6+\delta}$ and Tl$_2$Ba$_2$CuO$_{6+\delta}$, however, the longitudinal MR is found to have a similar magnitude as the transverse MR \cite{Ayres-2021}, behaviour that is inconsistent with conventional (Boltzmann) transport theory. Fig.~\ref{SI:Hparallel} compares the transverse and longitudinal MR for three overdoped LCCO films -- $x$ = 0.159, 0.174 and 0.188 -- at $T$ = 75 K. Clearly, the longitudinal MR is negligible at this temperature, implying that the MR of LCCO is entirely orbitally-driven. (At lower temperatures, a non-zero MR appears that we attribute to the emergence of superconducting fluctuations.) This anisotropy in the MR in electron-doped cuprates has been observed previously \cite{Dagan-PRL-2005, Jin-PRB-2009, Li-SciAdv-2019}, in addition to an isotropic \textit{negative} component in the longitudinal MR in PCCO below $x_{\rm FSR}$ that was attributed to AFM order \cite{Dagan-PRL-2005}.

\section*{Boltzmann formalism in electron-doped cuprates}

The purely orbital nature of the MR in OD LCCO motivates us to examine its $T$- and $H$-dependence within a standard Boltzmann treatment, albeit one modified to take into account the presence of a marked in-plane anisotropy of the scattering rate. We use the Shockley-Chambers Tube Integral Formalism (SCTIF) applied to a 2D metal in the time basis to avoid divergences that can occur as $B\rightarrow 0$. The SCTIF is the general solution of the Boltzmann transport theory equation within the relaxation-time approximation (RTA) and is used to determine the response of quasiparticles to an external electromagnetic field at arbitrary field strengths. The expression for the conductivity within the time basis, after neglecting the temperature broadening of the FS, is \cite{Shockley-PhysRev-1950, Chambers-1952}:

\begin{equation}
    \label{ch-LCCO-eq:SCTIF}
        \sigma_{ij} = \frac{1}{4\pi^3}\int_{FS}d^2k\frac{1}{\hbar v_F}qv_i\int_0^{\infty}qv_j(-t)P(t)dt,
\end{equation}

where $v_{i,j}$ are the velocity components with $\{i,j\} = \{x,y\}$, $P_{\phi}(t)$ is the probability of a quasiparticle starting at angle $\phi$ to survive for time $t$, defined as:

\begin{equation}
    \label{ch-LCCO-eq:prop}
       P_{\phi}(t) := \exp\left[ -\int_0^t\frac{dt'}{\tau(t')} \right] ,
\end{equation}

where $\tau(t)$ is the velocity relaxation time, and all other parameters take their usual values. The key parameter which determines the $H$-dependence of the MR is the mean-free-path $\ell (= v_F\tau)$ and in particular its anisotropy. The FS of LCCO does not exhibit sharp corners and is not in close proximity to a van Hove singularity. Thus, the $k$-dependence of $v_F$ (which incorporates the variation of the density of states around the FS) is anticipated to be limited. Indeed for $x=0.188$ where the FS curvature is most pronounced, $v_F$ varies by $\leq 10 \%$ \cite{Tang_PRB_2021}. It is thus the anisotropy of $\tau$ which is expected to dominate the MR. A standard tensor inversion is used to convert the magnetoconductivity to magnetoresistance.

Given the strong magnetic order observed below $x_{\rm FSR}$, spin correlations are likely manifest at all $x$ even when their obvious signatures in the MR vanish. Such spin correlations may necessitate going beyond the RTA by, for example, including the effects of current vertex corrections (CVCs), as highlighted in the main article. These CVCs have been predicted to give rise to a non-FL $T$-dependence of the resistivity and have a crucial impact on the Hall effect \cite{Kontani-RepProgPhys-2008}, as will be discussed in a later section. 

\section*{Tight-binding parameterisation}
    
The general tight-binding model for 2D systems with tetragonal symmetry is given by
        
\begin{equation}
    \begin{split}
        \label{Tight_binding:LCCO}
        E({\bf{k}}) = &\varepsilon_0 - 2t\left(\cos k_x a+\cos k_y a\right) -4t' \cos k_x a \cos k_y a \\
        &-2t'' \left(\cos 2k_x a+\cos 2k_y a\right) 
    \end{split}
\end{equation}        

where $t, \ t'$ and $t''$ are the nearest, next-nearest, and next-next nearest neighbour hopping parameters, and $\varepsilon_0$ is a measure of the chemical potential. We use the conditions given in Ref.~\cite{Tang_PRB_2021} where $t=0.3$ and $t''/t'=-0.5$ remain constant at all $x$, and we assume that the Luttinger count $x=\frac{2A_{\rm{FS}}}{A_{\rm{BZ}}}-1$ is satisfied in order to fix $\varepsilon_0$ and $t'/t$. Here, $A_{\rm{FS(BZ)}}$ is the area of the FS (Brillouin Zone), respectively. The TB parameters for LCCO are listed in Table \ref{tab:LCCO-tb}. LCCO is tetragonal with lattice parameters $a=4.01$  \AA \ and $c=12.38$ \AA. The standard $\hbar\vec{v} = \nabla_k E$ relation is used to obtain the components of the velocity in Eq.~\ref{ch-LCCO-eq:SCTIF}, leaving $\tau$ as the only parameter that needs to be explicitly defined.

\begin{table}[!h]
    \centering
    \begin{tabular}{|c|c|c|c|c|}
    \hline
    $x$   & $\varepsilon_0$    & $t$ & $t'/t$ & $t''/t'$ \\ \hline
    0.1   & -0.045   & 0.3 & 0.22   & -0.5     \\ \hline
    0.125 & -0.02059 & 0.3 & 0.225  & -0.5     \\ \hline
    0.15  & 0        & 0.3 & 0.23   & -0.5     \\ \hline
    0.175 & 0.033    & 0.3 & 0.235  & -0.5     \\ \hline
    0.2   & 0.0615   & 0.3 & 0.24   & -0.5     \\ \hline
    \end{tabular}
    \captionsetup{width=\textwidth}
    \caption{\label{tab:LCCO-tb} Tight-binding parameters for LCCO based on the requirements listed in Tang {\it et al.} \cite{Tang_PRB_2021}. Using $t=0.3$ and $t''/t'=-0.5$, $\varepsilon_0$ and $t'/t$ are found for $x=0.1-0.2$ through the Luttinger count.}
\end{table}

\section*{Simulations}

While the form of the scattering rate $\tau^{-1}(\varphi,T,x)$ (redefined as $\Gamma(\varphi,T,x)$ hereafter) defined in Eq.~(1) of the main article is a result of extensive modelling, it contains only those components required to capture all the essential features of the experimental resistivity and MR data (outside of the FSR regime) as a function of magnetic field strength, temperature and doping with a single doping-dependent variable parameter $g(x)$. In this section, we justify the inclusion of each component in $\Gamma(\varphi,T,x)$ and present alternative scenarios that capture some, but not all, of the key features of the MR.

\begin{figure*}
    \centering
    \includegraphics[width = 1\textwidth]{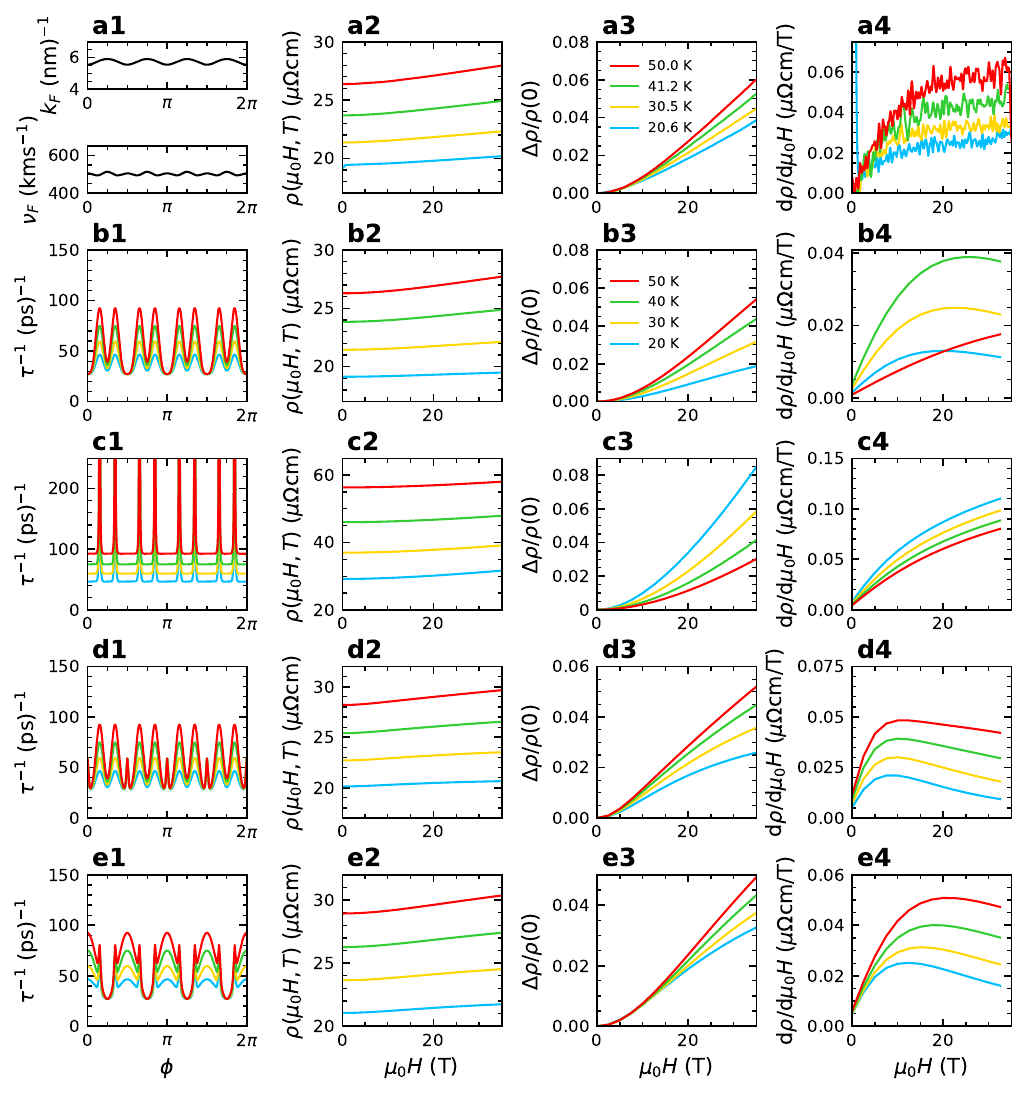}
    \caption{\textbf{(Inferior) simulations with alternative scattering rates for $x=0.159$}. ({\bf a}) Reproduction of row a) from Fig.~4 of the main article, for comparison. ({\bf b}) Anisotropic scattering (peaked at the AFM Brillouin zone boundary) without hotspots. While the zero-field resistivity is in excellent agreement with the experimental data, from panel b3 it is clear that the $T$-dependence of the MR at high magnetic fields cannot simultaneously be captured without the inclusion of hotspots. Furthermore, from the derivatives in panel b4, the simulated data at 50 K is approximately quadratic in field at all field strengths, whereas the data show a strong linear component. ({\bf c}) Isotropic inelastic scattering with stronger hotspot scattering ($10^5 \ \rm{ps}^{-1}$). In this case, the $T$-dependence of the MR is inverted highlighting the need for the anisotropic component in $\Gamma(\varphi,T,x)$. The stronger hotspot also amplifies the $T$-dependence of the zero-field resistivity in panel c2) beyond what is observed experimentally. ({\bf d}) Same as \textbf{b}) but with (small) hotspots positioned at the antinodes. In this case, the anisotropic components are spread evenly across the FS. As a result, the anisotropy is washed out rapidly with increasing $H$ and the MR tends towards saturation at all $T$. ({\bf e}) Anisotropic scattering peaked at the antinodes combined with hotspots at the AFM zone boundary. Again, the MR has a stronger tendency to saturate at all $T$.}
    \label{fig-SI:-sims}
\end{figure*}

The core element of $\Gamma(\varphi,T,x)$ is the 4-fold symmetric, double-peaked component that gives rise to 8 broad peaks, each located at the intersection of the underlying Fermi surface and the  AFM BZ boundary, as depicted in Fig.~3 of the main article as well as in panel b1) of Figure \ref{fig-SI:-sims}. For simplicity, both the $T$-linear and $T^2$ components of $\Gamma(\varphi,T,x)$ are assigned this anisotropic form, though only the $T$-linear term incorporates $g(x)$, as motivated by Fig.~6e). Combined with an isotropic impurity scattering rate $\Gamma_0$ of appropriate magnitude, this $T$-dependent scattering rate (labelled hereafter $\Gamma_{\rm T}$) captures the most striking feature of the data, namely the growth of $\Delta\rho/\rho(0)$ with increasing $T$ (Fig.~\ref{fig-SI:-sims}b3)). Recall that Kohler's rule would predict an opposite $T$-dependence for $\Delta\rho/\rho(0)$. The $H$-dependence of the MR is also reasonably well captured by $\Gamma_0 + \Gamma_{\rm T}$, bar two important details. Firstly, at low fields, the derivatives fan out from $H$ = 0 (Fig.~\ref{fig-SI:-sims}b4), while in the data (Fig.~\ref{fig-SI:-sims}a4), the initial slopes are $T$-independent. Secondly, the field dependence tends towards saturation as the anisotropy in $\Gamma_{\rm T}$ becomes washed out by ever-increasing cyclotron orbits at low $T$, and a purely quadratic MR at 50 K. In reality, the MR of overdoped LCCO remains $H$-linear up to fields as high as 65 T \cite{Sarkar-sciadv-2019}. This signifies the presence of loci (\lq hotpots') on the FS that impede cyclotron motion \cite{Hinlopen-PRR-2022} at all temperatures in this range.

It is important to realize that hotspots alone are insufficient to reproduce the experimental data. Row (c) of Figure \ref{fig-SI:-sims} shows the results of a simulation in which the anisotropy in $\Gamma(\phi)$ is contained solely in the strong hotspots positioned at the AFM zone boundary. While the $T$-dependence of $\rho(0,T)$ is reasonably well matched in this scenario, (Fig.~\ref{fig-SI:-sims}c2), the $T$-dependence of the MR is inverted (Fig.~\ref{fig-SI:-sims}c3). This demonstrates the clear need for the inclusion of $\Gamma_{\rm T}$ in the full expression for $\Gamma(\varphi,T,x)$.

The remaining two rows of panels in Figure \ref{fig-SI:-sims} consider alternative scenarios for $\Gamma(\varphi,T,x)$. According to a recent ARPES study, the nodal regions of the FS in LCCO exhibit non-FL behaviour \cite{Tang-npj-2022}, likely due to their close proximity to the AFM zone boundary. There are no known wave-vectors connecting two nodal points, however, and no experimental evidence to date for hotspot scattering at the nodes. Hence, we do not consider this scenario here. Instead, in row d) of Fig.~\ref{fig-SI:-sims}, we add smaller hotspots at the antinodal regions. This situation could arise, for example, from additional scattering off charge fluctuations between antinodes (grey double-headed arrows in Fig.~3 of the main article) \cite{daSilvaNeto-Science-2015,daSilvaNeto-SciAdv-2016}. 

\begin{table}[]
\begin{tabular}{|c|c|c|c|c|}
\hline
$\tau_{\rm imp}^{-1}$ & $\alpha_1 $       & $\alpha_2$                   & HS peak & HS width \\
(ps$^{-1}$)           & (ps$^{-1}$K$^{-1}$) & (ps$^{-1}$K$^{-2}$) &     (ps$^{-1}$)   &   \\ \hline
27                    & 0.75                 & 0.0112               &   50  &  0.05  \\ \hline
\end{tabular}
\captionsetup{width=\textwidth}
\caption{\textbf{Scattering rate parameters used to fit $\rho(H,T)$ of sample S1 at $x=0.159$}. $\tau_{\rm imp}^{-1}$ is the isotropic impurity scattering rate. $\alpha_1$ and $\alpha_2$ are the coefficients of the $T$-linear and $T^2$ resistivity, respectively. The hotspot is defined as a Gaussian with a fixed height and width. Note that all of these parameters are kept fixed for the simulations presented in the main article.}
\label{ED_Tab1)Scattering rate}
\end{table}

\begin{table}

\begin{longtable}{ | p{0.8cm} | p{0.8cm} | }
\hline
$x$   &  $g(x)$ $T = 40$ K \\ \hline
0.116 &  6.50  \\ \hline
0.119 &  5.75  \\ \hline
0.12  &  5.20  \\ \hline
0.125 &  3.85  \\ \hline
0.131 &  2.90  \\ \hline
0.134 &  2.52  \\ \hline
0.138 &  2.10  \\ \hline
0.141 &  1.87  \\ \hline
0.146 &  1.46  \\ \hline
0.149 &  1.45  \\ \hline
0.153 &  1.38  \\ \hline
0.156 &  1.18  \\ \hline
0.159 &  1.00  \\ \hline

\captionsetup{width=0.8\textwidth}
\caption{\textbf{Doping dependence of $g(x)$ for S1}. Here, the anisotropic $T$-linear scaling factor $g(x)$ is obtained by normalizing to the value at $x$ = 0.159 (where the simulation fitting of the $T$- and $H$-dependence of the MR was carried out) and varying $g(x)$ such that it simulated the zero-field resistivity and MR magnitude at $T$ = 40 K and $\mu_0H$ = 20 T of all other channels (without changing any other parameter). Accordingly, $g(x)$ is found to be largest at the lowest doping and drops monotonically across the series.}
\label{ED_Tab:x-dep}
\end{longtable}
\end{table}

\newpage

The resulting simulations show a MR which is quadratic at low fields but which rapidly transitions towards saturation, as can be seen in the 20 K MR curve in panel \ref{fig-SI:-sims}d3) and its derivative in panel \ref{fig-SI:-sims}d4). This tendency is due to the fact that the SCTIF concerns the average scattering rate around the FS and thus while this scenario contains a significant amount of anisotropy, its variation around the FS is effectively seen as an increase in isotropic scattering, resulting in a tendency towards MR saturation. 

Finally, in row e), we present a simulation based on dominant scattering from charge fluctuations, with sub-dominant hotspots at the AFM BZ boundary. In this case, the anisotropic inelastic component is a broad peak centred around the antinodes. While the initial slopes are essentially independent of temperature in this scenario (Fig.~\ref{fig-SI:-sims}e4), the MR again has a tendency towards saturation at higher field strengths, particularly at lower $T$.





\begin{table}

\begin{longtable}{ | p{0.8cm} | p{0.8cm} | }
\hline
$x$   & $g(x)$          \\ \hline
0.159 & 1.00            \\ \hline
0.165 & 0.85            \\ \hline
0.169 & 0.76            \\ \hline
0.174 & 0.66            \\ \hline
0.18  & 0.57            \\ \hline
0.184 & 0.55            \\ \hline
0.188 & 0.55            \\ \hline

\captionsetup{width=0.8\textwidth}
\caption{\textbf{Doping dependence of $g(x)$ for S2}.  Here, $g(x)$ is obtained from the coefficient of the $T$-linear resistivity shown in Data Fig.~6e) and normalized such that that the values of $g(x)$ in both films are the same at $x$ = 0.159. $g(x)$ is found to drop by roughly a factor of 2 across this doping range.}
\label{ED_Tab:x-dep_pt2}
\end{longtable}
\end{table}

The scattering rate parameters that provide the best fits to the $x$ = 0.159 data (shown in Fig.~4 of the main article) are listed in Table \ref{ED_Tab1)Scattering rate}. The hotspot scattering rate $\tau_{\rm HS}^{-1}$ is formed of a Gaussian curve whose peak and width are given in the last two columns. The HS peaks have a similar magnitude as the inelastic scattering rate, this results in an overall $\Gamma$ that is more peaked at the AFM Brillouin zone boundary, providing the anisotropy that is required to reproduce the MR at low $T$.

\begin{figure}
    \centering
    \includegraphics[scale=1]{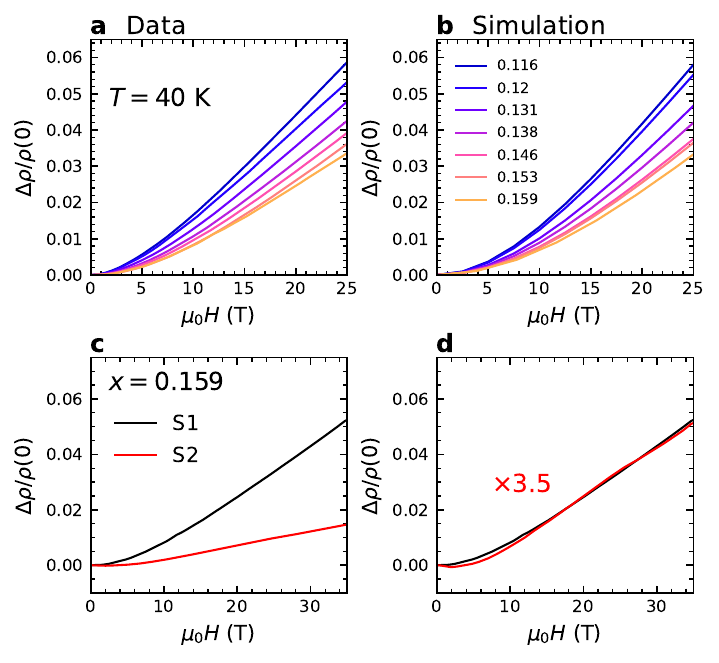}
    \caption{\textbf{Magnetoresistance and simulations for selected dopings from film S1}. ({\bf a}) The experimentally obtained $\Delta\rho/\rho(0)$ for 7 doping values on S1 at $T = 40$ K. ({\bf b}) The simulations for the same doping levels as in (a) where only $g(x)$ has been varied to capture the $x$-dependence. ({\bf c}) The unscaled and ({\bf d}) scaled MR of S2 at $x$=0.159. The origin of the reduced MR in S2 is most likely the negative MR component that is observable at low field strengths. This low-field negative MR contribution is seen in all channels in S2. As a result, we only use the zero-field resistivity, and more specifically, the evolution of the $T$-linear coefficient $A_1$, to deduce the value of $g(x)$ beyond $x$ = 0.159.}
    \label{SI:xDepSims}
\end{figure}

In order to model the $x$-dependence of the MR and $\rho(0,T)$, we assume that only the $T$-linear component of $\Gamma_{\rm T}$ changes with $x$, its variation defined by the dimensionless parameter $g(x)$ whose absolute values are listed in Table \ref{ED_Tab:x-dep}. Comparison of the raw $\Delta\rho/\rho(0)$ data and simulations for S1 at $T = 40$ K are shown in Fig.~\ref{SI:xDepSims}a) and b). We find that the evolution of both the $H$-dependence and magnitude of $\Delta\rho/\rho(0)$ with $x$ can be well described simply by varying $g(x)$. Due to the extent of $\Gamma_{\rm T}$ in momentum space, scaling $g(x)$ additionally affects the zero-field resistivity. The small discrepancy between the simulated and measured absolute values of the MR shown in Fig.~5b) of the main article could be removed by imposing fewer constraints on the scattering rate, though for simplicity, we have not done this. 

As $x$ increases, it is possible, of course, that the hotspots broaden or diminish in size, in tandem with $g(x)$, as one might expect upon approaching the FL regime. Indeed, we find that increasing the width of the hotspot by a factor of two or decreasing the size of the hotspot by a factor of 5 reduces the resistivity by 1 $\mu\Omega\mathrm{cm}$. This behaviour may explain the larger increase of $g(x)$ below $x = 0.14$ and at lower $T$: the sharpening and strengthening of hotspots would contribute to the increase in the resistivity in parallel with the anisotropic $T$-linear term. Note that the strength of the hotspot can only be quantified experimentally if breakdown in the form of MR saturation is observed. Thus, for the sake of transparency and simplicity, we keep the height of $\tau^{-1}_{\rm HS}$ constant at all $x$ and $T$.  The origin and nature of $\Gamma_{\rm T}$ will be discussed in more detail below. As mentioned in the main article, the MR of each voltage pair in S2 was found to contain a negative component, visible in Fig.~\ref{SI:xDepSims}c) and d) for the raw and scaled data, respectively. This made it difficult to model the MR for $x > 0.159$ without introducing an additional scaling factor. For this reason, $g(x)$ for $x > 0.159$ was taken directly from the evolution of the $T$-linear coefficient of $\rho(0,T)$ (i.e. the extrapolated $T = 0$ intercept of Fig. 6e), having first scaled the value of $g(x)$ for $x$ = 0.159 for S2 to that obtained in S1.

\section*{Incorporation of CVCs into the Boltzmann formalism}

\begin{figure*}
    \centering
    \includegraphics[width = 0.9\textwidth]{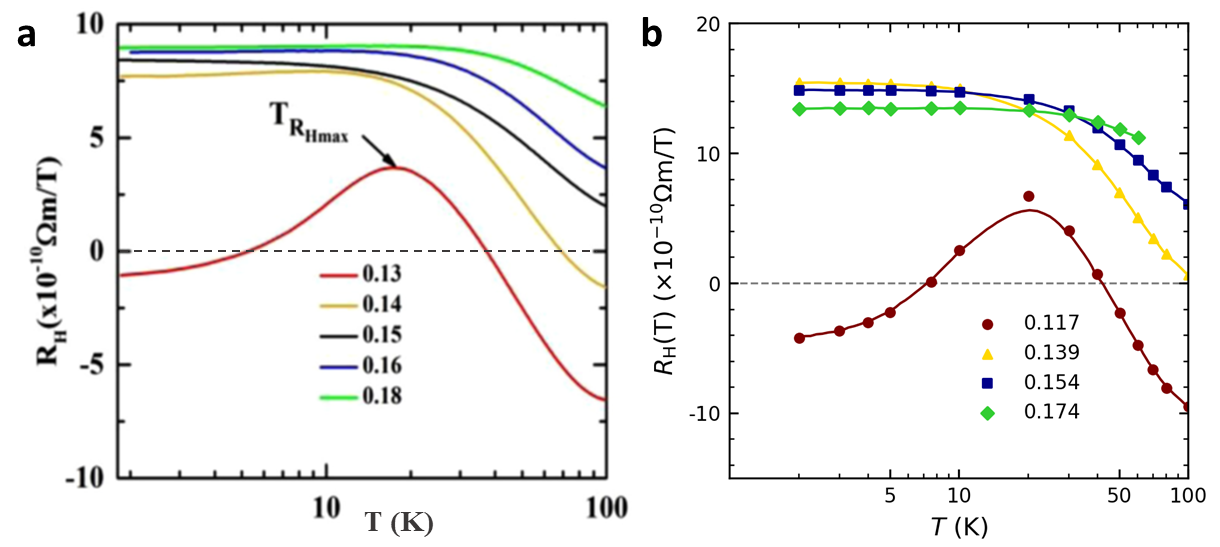}
    \caption{{\bf Hall coefficient in LCCO}. \textbf{(a)} $R_{\rm H}(T)$ curves in non-combinatorial thin films of LCCO over the doping range 0.13 $\leq x \leq$ 0.18. Reproduced from Ref.~\cite{Sarkar-PRB-2017}. \textbf{(b)} Selection of $R_{\rm H}(T)$ curves in combinatorial LCCO films over a similar doping range \cite{Yuan-Nature-2022}. Note the propensity of $R_{\rm H}(T)$ to decrease with increasing $T$ and tend towards negative values even for overdoped samples beyond $x_{\rm FSR}$ = 0.14. Both sets of data were obtained in a field of 14 T.}
    \label{fig:5-RH}
\end{figure*}

In a 2D metal, the Hall conductivity $\sigma_{xy}$ is determined by the \lq Stokes' area swept out by the $\vec{\ell}(\varphi)$ vector as $\varphi$ varies around the FS \cite{Ong-1991}. The FS of LCCO, as sketched in Figures 3 and 4 of the main article, has only hole-like curvature, meaning that $\vec{v}_{\rm F}$ has the same circulation at all $\varphi$. In this circumstance, even the inclusion of strong (orders-of-magnitude) anisotropy in $\tau^{-1}(\varphi)$ cannot cause $\sigma_{xy}$ and thereby $R_{\rm H}$ to become negative. In LCCO for $x > x_{\rm FSR}$, however, $R_{\rm H}(T)$ drops monotonically from close to its isotropic (Drude) value towards zero and even becomes negative at elevated temperatures \cite{Yuan-Nature-2022} (Fig.~\ref{fig:5-RH}). While a two-band model is able to account for the behaviour of the Hall resistance, it does not respect the FS measured in ARPES nor can it simultaneously account for the MR \cite{Li-PRL-2007}.  In the absence of any additional FSR, for which there is no evidence, the only effective way of explaining the sign change in $R_{\rm H}(T)$ is a breakdown of the RTA and the incorporation of current-vertex corrections (CVCs) into the calculation for $\sigma_{xy}$ \cite{Jenkins-PRB-2010}.
 
In this section, we develop a semi-quantitative picture of the evolution of $R_{\rm H}(T)$ by incorporating the effect of CVCs into the analysis of the Hall response. For these simulations, we use the Jones-Zener expansion. While this form of the Boltzmann equation is only applicable at low magnetic fields, $R_{\rm H}$ itself has only been measured to date at relatively low field strengths. Moreover, this form allows the use of analytical expressions for most of the key quantities.

We start by introducing analytical expressions for $k_F(\varphi)$ and $v_F(\varphi)$ that simplify the TB parameterizations above while preserving their four-fold symmetry:

\begin{equation}
        \label{eq:kF_LCCO}
        k_F(\varphi) = k_{F0}(1 + \alpha \sin^2(2\varphi))
\end{equation}        

\begin{equation}
        \label{eq:vF_LCCO}
        v_F(\varphi) = v_{F0}(1 + \beta \sin^2(2\varphi))
\end{equation}        

Here $k_{F0}$ = 5.424 $\times$ 10$^9$ m$^{-1}$, $\alpha$ = 0.11, $v_{F0}$ = 4.946 $\times$ 10$^5$ ms$^{-1}$ and $\beta$ = 0.03. Note that in our simulations of the Hall response, the variation in $v_F(\varphi)$ at the anti-nodes (ANs) has been flattened slightly relative to Eq.~(4) in order to approximate the TB parameterization more closely. This has no impact on $\rho(H, T)$. $v_x(\varphi)$  and $v_y(\varphi)$  are then obtained from $v_F(\varphi)$ using the analytical expressions:

\begin{equation}
    \begin{split}
        \label{eq:vx_LCCO}
        v_x(\varphi) = v_F(\varphi) \cos(\varphi - \gamma) \\  v_y(\varphi) = v_F(\varphi) \sin(\varphi - \gamma)
    \end{split} 
\end{equation}      
where \cite{Hussey-2003}
\begin{equation}
        \label{eq:gamma}
        \gamma(\varphi) = \tan^{-1} \left ( \partial[{\rm ln} (k_F(\varphi))]/\partial \varphi \right )
\end{equation}   

In electron-doped cuprates close to commensurate AFM order, the dominant scattering vector $\vec{Q}$ = ($\pi$, $\pi$). When CVCs are introduced, momenta linked through $\vec{Q}$ are heavily mixed, as a result of which the total current $J_k$ becomes approximately parallel to $v_k + v_{k+Q}$ and is thus no longer perpendicular to the FS. Previously, Kontani used linear response theory to derive the following expression to describe how $\vec{J}_k$ transforms in proximity to the AFM hotspot \cite{Kontani-RepProgPhys-2008}:

\begin{equation}
    \label{eq:HS_CVC}
        \vec{J}_k = \frac{1}{1 - \epsilon_k^2} (\vec{v}_k + \epsilon_k \vec{v}_{k\pm Q})
\end{equation}

where $\epsilon_k \approx (1 - c \xi_{\rm AFM}^2) < 1$, $\xi_{\rm AFM}$ is the AFM correlation length and $c$ is a constant. In order to incorporate CVCs into our Boltzmann framework, we first substitute this expression for $\vec{J}_k$ ($\vec{v}_k$) into the kinetic equation and recalculate $v_x$, $v_y$:

\begin{equation}
    \begin{split}
        \label{eq:vx_CVC}
        v_x^{\ast}(\varphi) = \frac{1}{1 - \epsilon(\varphi)^2} (v_x(\varphi) + \epsilon(\varphi) v_x(\varphi'))
        \\
        v_y^{\ast}(\varphi) = \frac{1}{1 - \epsilon(\varphi)^2} (v_y(\varphi) + \epsilon(\varphi) v_y(\varphi'))
    \end{split}
\end{equation}

where $\varphi'$ = $\varphi$ - ($\pi$, $\pi$). This process effectively links the hot-spots through the AFM $Q$-vector. For each value of $x$, the angles $\varphi_n$ associated with the hotspots are determined by the locations where the FS crosses the AFM zone boundary (see, e.g.~Fig.~\ref{fig:4-velocity}c)/d)). For $\epsilon(\varphi)$, we assume a Gaussian with a fixed breadth and 	height (but, for simplicity, ignore the prefactor  1/(1 - $\epsilon^2$)), using the expression:

\begin{equation}
    \label{eq:epsilon_CVC}
        \epsilon(\varphi) = \exp \frac{-(\varphi-\varphi_n)^2}{2\sigma^2} 
\end{equation}

with $\sigma$ = 0.07 (chosen to reduce overlap between adjacent hotspots). We then substitute these expressions for $v_x^{\ast}$ and $v_y^{\ast}$ into the Jones-Zener expansions for $\sigma_{xx}$ and $\sigma_{xy}$:

\begin{equation}
    \label{eq:sigma_xx}
        \sigma_{xx} = \frac{e^2}{2\pi^2 \hbar d} \int_{0}^{2\pi} \frac{k_F(\varphi) v_x^{\ast}(\varphi)\cos(\varphi-\gamma)}{\Gamma(\varphi)\cos\gamma} \,d\varphi 
\end{equation}

\begin{equation}
    \label{eq:sigma_xy}
        \sigma_{xy} = \frac{-e^3}{2\pi^2 \hbar^2 d} \int_{0}^{2\pi} \frac{v_x^{\ast}(\varphi)}{\Gamma(\varphi)} \frac{\partial [v_y^{\ast}(\varphi)/\Gamma(\varphi)]}{\partial \varphi} \,d\varphi 
\end{equation}

where $d$ is the $c$-axis lattice spacing. Finally, for $\Gamma(\varphi)$, we adopt the same form as given in Eq.~(1) of the main article which is reproduced here combined with the (isotropic) impurity scattering rate $\Gamma_0$:

\begin{figure*}
    \centering
    \includegraphics[width = 0.85\textwidth]{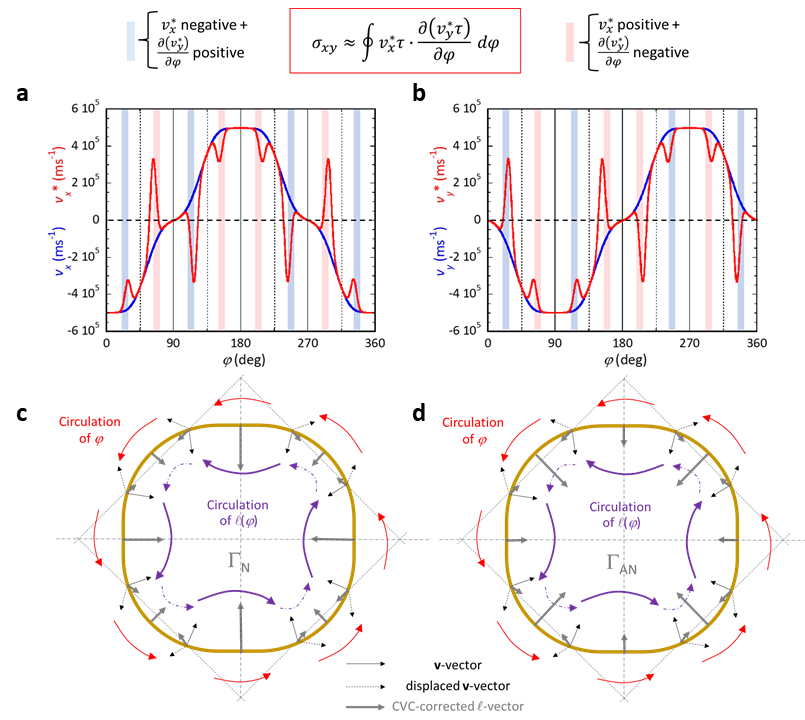}
    \caption{{\bf Possible influence of current vertex corrections (CVCs) on the Hall response in LCCO}. \textbf{(a)} Variation of $v_x(\varphi)$ (blue line) and its CVC-modified counterpart $v_x^{\ast}(\varphi)$ (red line). \textbf{(b)} Corresponding plot of $v_y(\varphi)$ and $v_y^{\ast}(\varphi)$. The shaded vertical bars represent those sections of the FS where the product of $v_x^{\ast}(\varphi)$ and $\partial v_y^{\ast}/\partial \varphi$ is negative, as described by the expressions above the panels. \textbf{(c/d)} Comparison of the impact on $\sigma_{xy}$ of the change in circulation of $\ell(\varphi)$ due to CVCs for the case where \textbf{(c)} {$\ell$ ($\Gamma$) at the nodes is shorter (greater) than at the anti-nodes (a scenario labelled $\Gamma_{\rm N}$ in panel c)) and \textbf{(d)} vice-versa ($\Gamma_{\rm AN}$). The length of the grey arrows reflect the magnitude of $\vec{\ell}(\varphi)$ at the symmetry points}. Wherever the circulation of $\ell(\varphi)$ opposes that of $\varphi$ (i.e. is anti-clockwise), the contribution to $\sigma_{xy}$ from that section is negative. $\sigma_{xy}$ itself is dominated by those regions of the FS where the curvature is largest. For LCCO, the largest curvature occurs around the zone diagonals. In panel c), the scattering rate at the nodes is greater than at the anti-nodes and so the (positive) contribution to $\sigma_{xy}$ is reduced relative to the negative contribution from the more extended regions between the hotspots, leading to a reduction in $\sigma_{xy}$. Conversely, in d), $\sigma_{xy}$ becomes larger (more positive) as the anisotropic scattering rate grows. }
    \label{fig:4-velocity}
\end{figure*}

\begin{equation}
    \label{eq:LCCO_scattering}
        \Gamma(\varphi) =   \Gamma_0 + \Gamma_{\rm HS}(\varphi) +  (g(x)\alpha_1T + \alpha_2 T^2)\sin^{\nu}\left[2\left(\varphi \pm \left(\frac{\pi}{4}- HS\right)\right)\right]
\end{equation}

where $\Gamma_{\rm HS}$ is the additional sharp Gaussian contribution at each hotspot and $\nu$ = 12. 

\section*{The anisotropic scattering term in LCCO}

Panels a) and b) of Fig.~\ref{fig:4-velocity} show, respectively, $v_x(\varphi)$ and $v_y(\varphi)$ together with their CVC-modified counterparts $v_x^{\ast}(\varphi)$ and $v_y^{\ast}(\varphi)$ for the parameterisations defined in Eq.~(8/9). The expression for the Hall conductivity $\sigma_{xy}$ of a 2D metal \cite{Ong-1991} (Eq.~(13) and reproduced in the box at the top of the figure) is set by the product of $\ell_x(\varphi)$ and $\partial \ell_y(\varphi)/\partial \varphi$. In the absence of anisotropy in $\ell(\varphi)$ or any CVCs, $\ell_x . \partial \ell_y$ is always positive. Introducing the CVCs, however, leads to $\varphi$-ranges where $v_x^{\ast}(\varphi)$ and $\partial v_y^{\ast}$/$\partial \varphi$ have opposite sign, as indicated by the vertical shaded bars in both panels. Within the Ong construction, this sign change arises due to a change in circulation of the $\vec{\ell}$-vector around the FS. These regions of opposite circulation are highlighted as thick solid arrows in panels c) and d) of Fig.~\ref{fig:4-velocity}. Correspondingly, these regions can contribute a component to $\sigma_{xy}$ of opposite sign, even when the scattering time $\tau$ is isotropic. The width of these regions is set by the width of the \lq influence' of the CVCs. Incorporating anisotropy into $\tau(\varphi)$ (or equivalently $\Gamma(\varphi)$) will act to enhance or reduce this effect, depending on the form of $\Gamma(\varphi)$ itself.

For $R_{\rm H}(T)$ to drop in magnitude with increasing $T$ relative to its isotropic, low-$T$ value, we require that $\ell$ in those regions in which the circulation of $\ell(\varphi)$ is opposite to that of $\varphi$ to be longer -- a scenario we label $\Gamma_{\rm N}$ in Fig.~12c). Conversely, if $\ell$ is shorter at the anti-nodes, the opposite effect occurs and $\sigma_{xy}$ will become larger as the anisotropy in the total scattering rate grows. This scenario, labelled $\Gamma_{\rm AN}$, is shown in Fig.~12d). The former scenario ($\Gamma_{\rm N}$) is most likely to arise when spin fluctuation scattering is dominant on account of the fact that the nodes are notably closer to the AFM hotspots. $\Gamma_{\rm AN}(\varphi)$, on the other hand, shown in Fig.~\ref{fig:4-velocity}d), is more likely to be associated with small-$\vec {Q}$ charge fluctuations, or strong electron-phonon scattering, as suggested by recent ARPES measurements on LCCO that report the presence of a ubiquitous antinodal kink at a characteristic (doping independent) energy scale $\approx$ 45 meV. Its ubiquity points towards coupling to phonons \cite{Tang-npj-2022}. While this phonon-derived mode may be manifest in the single-particle response, its impact on the transport properties is likely to be negligible.

Inspection of panels c) and d) of Fig.~\ref{fig:4-velocity} provides a qualitative understanding of which form of $\Gamma(\varphi)$ will act to reduce $R_{\rm H}$ with increasing $T$.  As explained in the figure caption, for $\Gamma_{\rm N}(\varphi)$ ($\Gamma_{\rm AN}(\varphi)$), counter-circulation of $\vec{\ell}(\varphi)$ occurs in those regions where $\ell$ is longest (shortest). Hence, as anisotropy in $\Gamma_{\rm N}(\varphi)$ grows with increasing $T$, the negative contributions to $\sigma_{xy}$ will also grow, causing $R_{\rm H}(T)$ to drop. Conversely, were $\Gamma_{\rm AN}$ to be the appropriate form of the $T$-dependent anisotropic component, $R_{\rm H}(T)$ would increase with increasing $T$.

\begin{figure*}[!b]
    \centering
    \includegraphics[width = 1\textwidth]{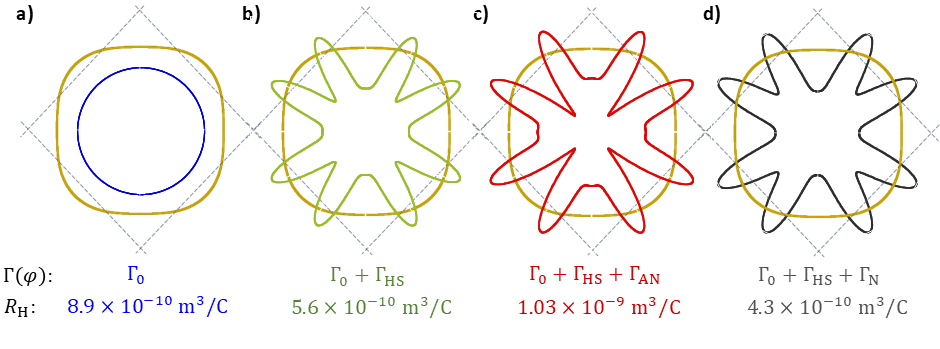}
    \caption{\textbf{Anisotropic scattering rates in LCCO}. Candidate forms of $\Gamma(\varphi)$ -- specified in the top line -- that are inserted into the Jones-Zener expansion of the Boltzmann transport equation to calculate the corresponding $R_{\rm H}$ -- specified in the second line below the plots. The forms of $\Gamma(\varphi)$ are: \textbf{(a)} $\Gamma_0$ (isotropic); \textbf{(b)} $\Gamma_0$ + $\Gamma_{\rm HS}$ (including a set of peaks associated with the AFM hotspots on top of an isotropic component); \textbf{(c)} $\Gamma_0$ + $\Gamma_{\rm HS}$ + $\Gamma_{\rm AN}$ (in which the broad anisotropic component is peaked at the anti-nodes); \textbf{(d)} $\Gamma_0$ + $\Gamma_{\rm HS}$ + $\Gamma_{\rm N}$ (in which the broad anisotropic component is peaked at along the zone diagonals).}
    \label{fig:5-Gamma}
\end{figure*}

This analysis is put on a more quantitative footing in Figure \ref{fig:5-Gamma}, where $R_{\rm H}$ is calculated within the same Jones-Zener formalism for different forms of $\Gamma(\varphi)$. For all panels, we use a parameterisation for $k_F$ and $v_F$ that is consistent with those used for $x$ = 0.159 in Fig.~4 of the main article. Fig.~\ref{fig:5-Gamma}a) shows the isotropic case; $\Gamma(\varphi)$ = $\Gamma_0$. In this instance, we obtain $R_{\rm H}$ = +8.7 $\times$ 10$^{-10}$ m$^3$/C, about 20\% higher than the Drude estimate for a (hole) carrier density of 1-0.159 = 0.841 per unit cell and in good agreement with the data shown in Fig.~\ref{fig:5-RH}) for $x$ = 0.16 at low $T$, even with the inclusion of CVCs. For $\Gamma(\varphi)$ = $\Gamma_0$ + $\Gamma_{\rm HS}$ (Fig.~\ref{fig:5-Gamma}(b)), we find $R_{\rm H}$ = +5.6~$\times$~10$^{-10}$~m$^3$/C, showing that the addition of hotspot scattering leads to a slight reduction in $R_{\rm H}$.

As argued above, a larger $R_{\rm H}$ is found when the scattering rate is larger in the anti-nodes. Fig.~\ref{fig:5-Gamma}c) depicts a form of $\Gamma(\varphi)$ = $\Gamma_0$ + $\Gamma_{\rm HS}$ + $\Gamma_{\rm AN}$. This form of scattering is somewhat artificial, since the scattering rate is larger where the hotspots are separated by a greater distance, but could reflect a scenario in which strong charge fluctuations coexist with the AFM hotspots. In this case, $R_{\rm H}$ = +1.03 $\times$ 10$^{-9}$ m$^3$/C, confirming that additional scattering at the anti-nodes leads to a further increase in $R_{\rm H}$ in the presence of CVCs. Hence, as anisotropy in $\Gamma(\varphi)$ grows with increasing $T$, so will $R_{\rm H}$. This is in the opposite sense to what is observed experimentally \cite{Sarkar-PRB-2017, Yuan-Nature-2022}. Clearly, in order to simulate a vanishing $R_{\rm H}$, the anisotropic component of $\Gamma(\varphi)$ needs to be inverted.   

\begin{figure*}
    \centering
    \includegraphics[width = 0.75\textwidth]{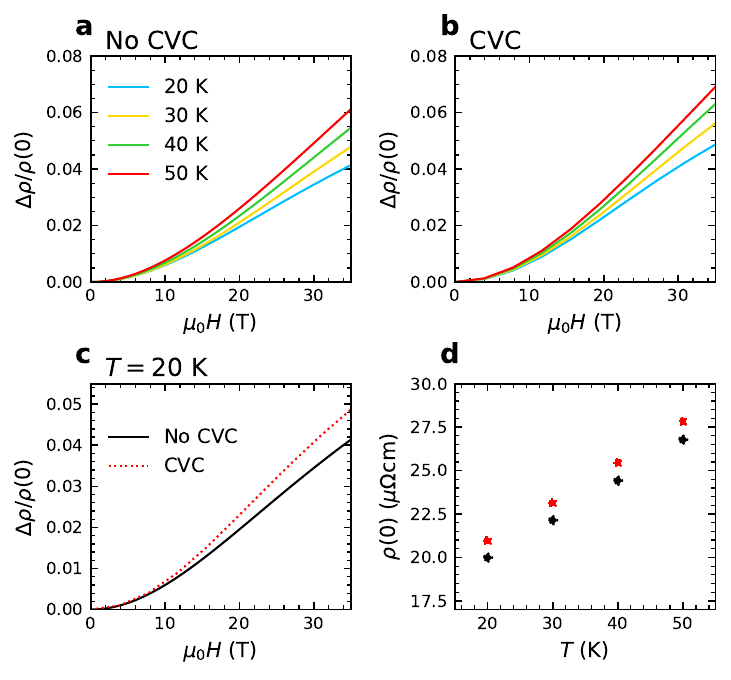}
    \caption{\textbf{Effect of current vertex corrections on the MR and zero-field resistivity.} \textbf{(a)}  The MR for$ x = 0.159$ without using CVCs, as in the main article. \textbf{(b)} The same with CVCs. \textbf{(c)} The MR at 40 K without (black) and with (red) CVCs. \textbf{(d)} 
    The zero-field resistivity without (black) and with (red) CVCs. Incorporating CVCs into the simulations of the MR has almost no effect on the magnitude and form of the MR, nor does it significantly impact the zero-field resistivity. 
    }
    \label{fig:CVC-comparison}
\end{figure*}

This is done in Fig.~\ref{fig:5-Gamma}d) where now  $\Gamma(\varphi)$ = $\Gamma_0$ + $\Gamma_{\rm HS}$ + $\Gamma_{\rm N}$. Using the same parameters that were used to model the MR in the main article (in the absence of CVCs), we find $R_{\rm H}$ = +4.3 $\times$ 10$^{-10}$ m$^3$/C, i.e. about half the value found for $\Gamma_0$ only. Moreover, as we show explicitly in Fig.~\ref{fig:CVC-comparison}, the incorporation of the modified velocities into the SCTIF has a negligible effect on the modeling of the MR. From panels a) and  b), it is clear that the effect on the MR is small; this is again evident in panel c) where the 20 K curves from both simulation are overlaid. Panel d) also demonstrates only a modest difference in $\rho(0,T)$. The effect of CVCs on the MR is small since (a) the MR is not influenced by the circulation of $\ell$ around the FS (in contrast to the Hall conductivity), and (b) the MR is determined largely by those sections of FS with the longest $\ell$, i.e. away from the hotspots. At the hotspots themselves, $\ell$ is extremely short and thus, those regions provide a negligible contribution to the total (magneto)-conductivity, even if their presence does have a strong influence on the $H$-dependence of the MR (through impeded cyclotron motion). Hence, the inclusion of CVCs will not affect the resistivity or the low-field MR in a significant way.


In summary, we find that the inclusion of $\Gamma_{\rm N}$ of a magnitude that leads to a doubling of the MR (Fig.~4 of the main article) also causes $R_{\rm H}$ to halve. This form of $\Gamma(\varphi)$ thus enables the two key $T$-dependences in the magnetotransport of LCCO beyond $x > x_{\rm FSR}$ to be simulated within the same parameterisation. This result also echoes an earlier study of the frequency ($\omega$) dependence of the Hall angle in overdoped PCCO where a sign change of $R_{\rm H}(T, \omega)$ in a FS with only singular curvature was also interpreted as evidence for AFM CVCs \cite{Jenkins-PRB-2010}.

This suggests that it should be possible to reproduce both $R_{\rm H}(T,x)$ and the MR from the ARPES-derived FS of LCCO using a single set of parameters. Given the large number of variables (width, height of hotspot, precise form of the anisotropic component, etc…), however, here we simply present this semi-quantitative analysis to justify the phenomenological form of $\Gamma(\varphi)$ and close this section with some preliminary thoughts on its possible physical origin.

The phenomenological form of $\Gamma(\varphi)$ -- necessitated primarily by the unusual growth in $\Delta\rho/\rho(0)$ in LCCO over a temperature range in which $\ell$ itself is diminishing -- has three principal features: $T$-linearity, large, smooth anisotropy and maxima at the AFM BZ boundary. The general approach to account for such a combination of features is to invoke the presence of low-lying bosonic excitations. The most natural candidates for these excitations are magnon modes in the vicinity of $\vec{Q} = (\pi,\pi)$. As shown in Fig.~3 of the main article, the node is three times closer to the AFM zone boundary than the anti-node. Thus, any spectral broadening of the $(\pi,\pi)$ mode (due to magnon lifetime effects) or any thermal occupation of modes up to about 10~meV at 50~K will disproportionately widen the hotspot towards the node. Both effects have been observed in electron-doped cuprates \cite{Yamada_1999, Wilson-2006, Fujita-2006} and thus offer a viable explanation as to why $R_{\rm H}(T)$ decrease with increasing $T$ while the MR increase (at least up to 70 K). This also may explain the $T$-linearity of $\Gamma(\varphi)$.

The final key question to address, though, is how scattering off these magnon modes can induce such a large anisotropy in $\Gamma$. Indeed, recent calculations based on the dynamical cluster approximation applied to overdoped cuprates show that scattering off low-lying spin excitations generates only mild anisotropy in the scattering rate \cite{Wu-PNAS-2022}. One possible source for the enhancement seen in experiment is small-angle scattering. Cuprates are known to exhibit primarily out-of-plane disorder in the charge-transfer layers rather than in the CuO$_2$ planes, resulting in small-angle scattering on the electrons \cite{Varma-2001, Abrahams-2003}. Small-angle scattering can be understood as Brownian motion or an effective diffusion process across the Fermi surface, and the expected time taken to diffuse into the nearby hotspots scales as the sector size squared (where the hotspots act as `sinks' similar to the necks in elemental Cu) \cite{Pippard-1964}. If the hotspot in LCCO indeed widens predominantly towards the node with increasing temperature, then the shrinking nodal sector can cause a marked increase in the efficacy of small-angle scattering to relax the velocity, thereby enhancing the anisotropy between the nodal and anti-nodal region. Thus, one possible interpretation of the increasing anisotropy in LCCO with temperature through the $g(x)$ term is that small-angle scattering exacerbates the anisotropic scattering rate. Although modeling this scenario is beyond the present work, we suggest hotspot broadening and small-angle scattering may explain the origin of the unusual $T$-dependent anisotropy in $\Gamma$ obtained here.

\section*{Inclusion of inelastic processes within Boltzmann}

Within the Boltzmann approach, quasiparticles travel on their cyclotron orbits without changing their energy and without decaying into other particles. In other words, only the momentum (not the energy) of the quasiparticle decays. Nevertheless, Boltzmann theory, specfically within the  relaxation time approximation (RTA), has been used to model the resistivity of metals effectively for over a century. In reality, Boltzmann transport within the RTA cannot differentiate between elastic and inelastic scattering. Boltzmann cares only about the relaxation of \lq velocity’ (through the vector $\ell$) – no matter if energy is exchanged in the process - just like it cannot distinguish between phonons, magnons, impurities or quasi-particle lifetimes from imaginary self-energies. In some sense, this is a disadvantage as it provides no direct information (we infer spin here because the hot spots are motivated by finite-$Q$ spin waves and because the inclusion of CVCs associated with these spin waves enables the $T$-dependence of the Hall response to be also captured, and we infer inelastic because $T$-activated phenomena naturally involve energetics). At the same time, it is also an advantage in the sense that Boltzmann is agnostic, because otherwise we would have even more fitting parameters and procedures. How momentum exactly relaxes in dc transport and if this relaxation involves a Planckian bound is an important open question and, no matter if we treat copper or cuprates, remains an active field of research.

In his seminal work treating spin fluctuation scattering near a QCP in the presence of disorder, Rosch applied a Boltzmann approach incorporating transition rates associated with both elastic impurity scattering and inelastic scattering from spin fluctuations \cite{Rosch-1999}. Rosch argued that this approach works provided that the spin fluctuations are assumed to stay in equilibrium, an approximation that is valid if the spin fluctuations lose their momentum effectively by Umklapp or impurity scattering. These are the same conditions we are assuming here.

\section*{Evolution of high-field slope over the full ($T$, $x$) range}

\begin{figure*}[!h]
    \centering
    \includegraphics[width = 0.65\textwidth]{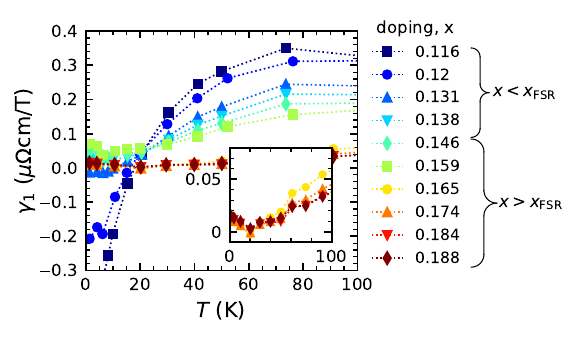}
    \caption{{\bf The link between the scattering rate anisotropy and AFM correlations}. $\gamma_1(={\rm d}\rho /{\rm d}\mu_0H|_{H = 33 \ \rm{T}}$) vs. $T$ for selected doping values across the range studied. For $x<x_{\rm FSR}$, $\gamma_1$ falls smoothly below 75 K and becoming negative below 20 K without any discontinuities, indicating that the drop in the MR is a consequence of AFM order and that correlations remain present at high temperatures. At higher $x$, $\gamma_1$ also falls below 75 K suggesting that AFM correlations still play a crucial role in the transport properties at elevated doping levels. The decrease in $\gamma_1$ with increased $x$ is then a reflection of the weakening AFM order when doped further away from the AFM phase.}
    \label{fig:6-slope}
\end{figure*}

As a further, final hint to the origin of the anisotropic scattering rate in LCCO, Fig.~\ref{fig:6-slope} shows the high-field slope $\gamma_1$, defined as the value of d$\rho$/d$\mu_0 H$ at 33 T, with respect to $T$ and $x$. At all carrier concentrations, irrespective of AFM order, the slope of the MR falls as $T<75$ K. Within the AFM regime, the slope becomes negative below $T\sim 20$~K without discontinuities suggesting that the origin of its $T$-dependence across the entire $T$-range is closely linked to the evolution of AFM correlations, given the negative slope is associated with AFM order \cite{Dagan-PRL-2005}. When $x>x_{\rm FSR}$, the slope remains positive at all $T$, but its magnitude also falls below 75~K before levelling off at 20 K. 

The smooth variation of the MR and the distinct similarity across the AFM boundary strongly links the unusual $T$-dependence of the MR below 75~K to AFM spin correlations which persist across the doping series. Note that the $T$-dependence diminishes as $x$ increases, reflecting as it does the weakening spin fluctuations implied by the decrease of the scaling parameter $g(x)$. Alternatively, the monotonic $x$-dependence is also evident from the gradual variation in d$\rho$/d$\mu_0H$ (plotted versus $H/T$ in Fig.~2h)-l) of the main article) with increasing $x$. The low-$T$ negative component seen at low $x$ also evolves continuously, becoming positive with no discontinuity near $x_{\rm FSR}$.

\bibliography{bibliography}

\end{document}